\documentclass[11pt]{article}
\usepackage{latexsym}
\usepackage{amsmath,amsfonts,amsbsy,amssymb,amscd}
\usepackage{fancybox,fancyhdr}
\usepackage{cite}
\usepackage{pstricks,pst-node,psfrag}
\usepackage{graphicx,epsfig,picture,graphics}
\usepackage{slashed}
\usepackage[all]{xy}
\usepackage{bbold}

\psset{unit=1.3cm,linewidth=.5pt,radius=.2}  % controls the size of diagrams

\usepackage{float}                          % to force floaters to stay Here
\usepackage{lscape}                         % landscape laid out pages
\usepackage{bm}
\usepackage{tikz}
\newcommand{\hoch}[1]{$\, ^{#1}$}
\def\rmi{{\rm i}}

\usepackage{color}
\definecolor{MyDarkBlue}{rgb}{0.15,0.15,0.45}    \definecolor{MyGreen}{rgb}{0.15,0.45,0.45}
\definecolor{MyPurple}{rgb}{0.55,0.25,0.55}
\usepackage[linktocpage=true]{hyperref}
\hypersetup{colorlinks=true,citecolor=MyGreen,linkcolor=MyPurple,urlcolor=MyGreen}

\begin{document}

%\begin{flushright}
% \vspace*{1cm}
%\hfill{ UG-15-68 \\ TUW-15-15}
%\end{flushright}
%\vskip 4cm

\begin{center}
{\Large \bf Critical $\mathcal{N}=(1,1)$ General Massive Supergravity}
\end{center}
\vspace{25pt}
\begin{center}
{\Large {\bf }}

\vspace{15pt}

{\Large Nihat Sadik Deger\hoch{1,2}, George Moutsopoulos\hoch{1} and Jan Rosseel\hoch{3}}

\vspace{30pt}

\hoch{1} {\it Department of Mathematics, Bogazici University,\\
Bebek, 34342, Istanbul, Turkey}
\vspace{4pt}

\hoch{2} {\it Feza Gursey Center for Physics and Mathematics,\\
Bogazici University, Kandilli, 34684, Istanbul, Turkey}\\

\vspace{5pt}

\texttt{sadik.deger@boun.edu.tr, gmoutso@gmail.com} \\

\vspace{10pt}

\hoch{3} {\it Faculty of Physics, University of Vienna,\\
Boltzmanngasse 5, A-1090, Vienna, Austria}\\

\vspace{5pt}

\texttt{jan.rosseel@univie.ac.at}

\vspace{45pt}
\underline{ABSTRACT}
\end{center}
In this paper we study the supermultiplet structure of $\mathcal{N}=(1,1)$ General Massive Supergravity at
non-critical and critical points of its parameter space. To do this, we first linearize the theory around its
maximally supersymmetric AdS$_3$ vacuum and obtain the full linearized Lagrangian including fermionic terms. 
At generic values, linearized modes can be organized as two massless and 2 massive multiplets where supersymmetry 
relates them in the standard way. At critical points logarithmic modes appear and we find that in 
three of such points some of the supersymmetry transformations are non-invertible in logarithmic multiplets. 
However, in the fourth critical point, there is a massive logarithmic multiplet with invertible supersymmetry 
transformations.

%\noindent

\vspace{15pt}

\thispagestyle{empty}

\vspace{15pt}

 \vfill

\voffset=-40pt

\newpage

\tableofcontents

% \addtocontents{toc}{\protect\setcounter{tocdepth}{2}}

% \newpage

% \thispagestyle{fancy}

%%%%%%%%%%%%%%%%%%%%%%%%%%%%%%%%%%%%%%%%%%%%%%%%%%%%%%%%%%%%%%%%%%%%%%%%%%%%%%

%%%%%%%%%%%%%%%%
%%%%%%%%%%%%%%%%                      Cut here
%%%%%%%%%%%%%%%%

\section{Introduction}

Logarithmic conformal field theories (log CFTs) \cite{Gurarie:1993xq} are 
conformal field theories in 
which the Hamiltonian is non-diagonalizable but where 
instead the operator spectrum decomposes in non-trivial Jordan cells. Certain 
operators are thus accompanied 
by so-called 
`logarithmic partner' operators, with which they form a Jordan cell. 
Log CFTs are typically non-unitary and hence they can be used 
to describe systems where non-unitarity is a 
feature, such as open quantum systems or systems with (quenched) disorder. 
Indeed, log CFTs have been considered in condensed matter physics and 
statistical mechanics in a variety of contexts, such as e.g. two-dimensional 
turbulence, critical polymers, abelian sandpile models, percolation, the 
fractional quantum Hall effect and systems with (quenched) disorder (see e.g. 
\cite{Flohr:1996ik,Grumiller:2013at} for references).

Jordan cells can also be found in gravitational theories 
in anti-de Sitter (AdS) space-times that include higher-derivative terms for 
the metric field. This was remarked for the first time in 
\cite{Grumiller:2008qz} in the context of so-called Topologically Massive 
Gravity (TMG), a three-dimensional gravity theory that includes a 
parity-violating three-derivative term for the metric. TMG is a specific 
instance of so-called `General Massive Gravity' (GMG) theories 
\cite{Bergshoeff:2009hq,Bergshoeff:2009aq}, 
that extend TMG with higher-derivative terms for the metric with up to four 
derivatives. Upon linearizing GMG theories, one ordinarily finds that their 
spectrum of 
linearized modes can be 
organized in eigenstates of the AdS Hamiltonian that correspond to 2 
massless 
graviton modes (that are pure gauge in three dimensions) and 2 massive graviton 
modes. For particular tunings of the GMG coupling constants however, one 
finds that some of the linearized modes become degenerate (e.g. some massive 
modes become massless) and that new so-called 
`logarithmic modes' appear in the spectrum. The spectrum of 
linearized modes 
then organizes itself in non-trivial Jordan cells of the AdS Hamiltonian. 
Points in the parameter space of GMG for which this phenomenon happens are 
often called `critical points' and GMG at such a point is then referred to as a `critical gravity' theory. The parameter space of GMG allows for 
several such points, where different modes become degenerate and different 
logarithmic modes appear. Critical gravity theories can also be found in higher dimensions with similar properties \cite{Lu:2011zk,Deser:2011xc,Bergshoeff:2011ri}. For other approaches to massive supergravity in four dimensions, see e.g. \cite{Gates:2013tka}.

The appearance of Jordan cells in critical gravity led to the conjecture that 
critical gravity theories can provide gravitational duals of log CFTs, where 
e.g. 
the energy momentum tensor acquires logarithmic partner operators. This 
 duality is often called the AdS/log CFT correspondence and  
it conjectures that the logarithmic gravity modes are dual to sources and vevs 
for
logarithmic partner operators in the CFT spectrum. The non-diagonalizability of 
the AdS Hamiltonian is thus translated 
into the non-diagonalizability of the CFT Hamiltonian. As there are various 
critical points in the parameter space of GMG, various versions of the 
AdS/log CFT correspondence have been proposed, within the context of GMG 
alone. The AdS/log CFT correspondence has been checked rather extensively 
during recent years. In particular, holographic techniques, such as holographic 
renormalization have been extended to take the presence of logarithmic modes 
into account and two- and three-point functions have been calculated on the 
gravity side and were found to be compatible with those of log CFTs. Similarly, 
one-loop partition functions of critical gravity theories were found to agree 
with the partition functions of log CFTs, to the extent that the latter are 
known. We refer to the review \cite{Grumiller:2013at} for details and 
references.

In contrast to the purely bosonic case, supersymmetric versions of the AdS/log CFT correspondence have not been studied much in the literature yet. 
In order to do that, a better understanding of the supermultiplet structure of the various linearized modes at a critical point of the dual 
gravity theory is required. This has been investigated in four-dimensional $\mathcal{N}=1$ critical supergravity models in \cite{Lu:2011mw}. For the purpose 
of studying the AdS/log CFT correspondence, the three-dimensional case is however more interesting, as the conjectured two-dimensional log CFTs are better 
understood than their higher-dimensional counterparts and the conjecture might thus be checked in more specific detail. Furthermore, the parameter space of three-dimensional critical gravity models is typically richer than that of higher-dimensional models. In this paper, we will therefore consider supersymmetric three-dimensional GMG models at various critical points and study how the various linearized modes form supermultiplets.

Various supersymmetric extensions of GMG exist in the literature. In particular, supersymmetric TMG has been constructed in \cite{Deser:1982sw} and $\mathcal{N}=1$ 
versions of full GMG have been constructed in \cite{Andringa:2009yc,Bergshoeff:2010mf,Bergshoeff:2010iy}. The bosonic Lagrangian and supersymmetry transformation 
rules of off-shell $\mathcal{N}=(2,0)$ and $\mathcal{N}=(1,1)$ GMG have been obtained using the method of superconformal tensor calculus in \cite{Alkac:2014hwa}, 
while a full superspace construction is given in\cite{Kuzenko:2011rd,Kuzenko:2013uya,Kuzenko:2015jda}. Some exact supersymmetric solutions of $\mathcal{N}=(2,0)$ and $\mathcal{N}=(1,1)$ GMG have 
been found in \cite{Deger:2013yla,Alkac:2015lma,Deger:2016vrn}. In this paper, we are going to consider the $\mathcal{N}=(1,1)$ supersymmetrization of GMG. We prefer to 
consider $\mathcal{N}=2$ models over $\mathcal{N}=1$ models, since the supermultiplet structure is richer. The $\mathcal{N}=(1,1)$ multiplet e.g. contains a vector, that 
is dynamical in the GMG theory. We do not consider the $\mathcal{N}=(2,0)$ models in this paper, since unlike $\mathcal{N}=(1,1)$ models these do not seem to have 
supersymmetric AdS vacua with ghost-free spectrum, when higher-derivative terms are present\cite{Alkac:2014hwa}. We will thus start from the $\mathcal{N}=(1,1)$ 
theories of \cite{Alkac:2014hwa} and perform a linearization of the theory around its maximally supersymmetric AdS vacuum, to study the spectrum of linearized modes and 
their linearized supersymmetry transformation rules. This will allow us to identify various critical points where logarithmic modes appear and to study the structure 
of the supermultiplets to which these logaritmic modes belong. In particular, we will show that there are four classes of critical points. One class is characterized 
by the fact that there is one supermultiplet containing logarithmic modes, along with massless and massive modes. A second class contains two such supermultiplets. A 
third class contains a supermultiplet with logarithmic and doubly logarithmic modes, along with massless and massive modes. Finally, the fourth class contains a 
supermultiplet of massive and logarithmic massive modes. We will devote special attention to the supersymmetry transformation rules that connect the various modes in the supermultiplets at critical points. In particular, we will see that for the first three classes of critical points, some of the supersymmetry transformations are not invertible. This is similar to what has been observed in four-dimensional $\mathcal{N}=1$ critical supergravity \cite{Lu:2011mw}.

The outline of this paper is as follows. In section \ref{sec:GMGlin}, we review the bosonic Lagrangian and off-shell supersymmetry transformation rules 
of $\mathcal{N}=(1,1)$ GMG. The fermionic terms of the Lagrangians of $\mathcal{N}=(1,1)$ GMG are not given in \cite{Alkac:2014hwa}. Since our analysis 
requires the linearized fermionic equations of motion, we will here also construct the full linearized Lagrangian, including fermionic terms, starting 
from the linearized bosonic Lagrangian and supersymmetry transformation rules. The spectrum of linearized modes, along with their linearized supersymmetry 
transformation rules for generic non-critical points in parameter space, is then studied in section \ref{sec:noncritspectrum}. These results are then used 
as a starting point for section \ref{sec:critpoints}, where the various critical points are discussed and the supermultiplets of logarithmic modes are identified 
and discussed. We end with conclusions and an outlook for future work in section \ref{sec:concl}. Finally, appendix A contains some useful notation and conventions.

\section{Linearized $\mathcal{N}=(1,1)$ General Massive Supergravity} \label{sec:GMGlin}

In this section, we will consider $\mathcal{N}=(1,1)$ General Massive Supergravity and its linearization around the maximally supersymmetric 
AdS$_3$ vacuum. Special attention will be given to the fermionic terms in the linearized action, that have not appeared in the literature before and 
that will be derived here by supersymmetrizing the linearized bosonic action.

\subsection{$\mathcal{N}=(1,1)$ General Massive Supergravity}

The field content of the off-shell $\mathcal{N}=(1,1)$ supergravity multiplet is given by the vielbein $e_\mu{}^a$, a vector field $V_\mu$, a complex scalar $S$ 
and two Majorana gravitini, that we will combine in a complex Dirac spinor $\Psi_\mu$. The off-shell supersymmetry transformation rules were derived in 
\cite{Rocek:1985bk,Nishino:1991sr,Cecotti:2010dg,Alkac:2014hwa}
\begin{align}
\delta e_\mu{}^{a}  &=  \frac12\bar{\epsilon}\gamma^{a}\Psi_{\mu}+ \mathrm{h.c.} \,, \nonumber \\
\delta\Psi_\mu  &=  D_{\mu}(\omega)\,\epsilon-\frac12 \rmi V_{\nu}\,\gamma^{\nu}\gamma_{\mu}\,\epsilon
-\frac12 S\gamma_\mu \left(B\epsilon\right)^{*}\,, \nonumber \\
\delta V_\mu  &=  \frac18 \rmi \bar{\epsilon}\,\gamma^{\nu\rho}\gamma_{\mu}\left(\Psi_{\nu\rho}-{\rmi} V_{\sigma}\gamma^{\sigma}\gamma_{\nu}\,\Psi_{\rho}-S\gamma_\nu \left(B\Psi_{\rho}\right)^{*}\right)+ \mathrm{h.c.} \,, \nonumber \\
\delta S  & =  -\frac14 \tilde{\epsilon}\,\gamma^{\mu\nu}\left(\Psi_{\mu\nu}
-\rmi V_{\sigma}\,\gamma^{\sigma}\gamma_{\mu}\Psi_{\nu}-S\gamma_{\mu}\left(B\Psi_{\nu}\right)^{*}\right)\,,
\label{eq:susyrulesnl}
\end{align}
where
\begin{equation}
  \label{eq:cdpsi}
  D_\mu(\omega) \epsilon = \left(\partial_\mu + \frac14 \omega_\mu{}^{ab} \gamma_{ab} \right) \epsilon \,, \qquad \Psi_{\mu\nu} = 2 D_{[\mu}(\omega) \Psi_{\nu]} \,.
\end{equation}
We refer to appendix \ref{app:conv} for our notation and conventions regarding complex spinors. 

The most general $\mathcal{N}=(1,1)$ supergravity action that includes up to four derivatives has been derived in \cite{Alkac:2014hwa}. This in particular 
includes an $\mathcal{N}=(1,1)$ supergravity version of General Massive Gravity, which will be the focus of this paper. The bosonic part of the Lagrangian 
is given by
\begin{align} \label{N11GMGnl}
e^{-1} \mathcal{L} &= \sigma \left( R + 2 V^2 - 2 |S|^2\right) + M A \nonumber \\
& \ \ \ -\frac{1}{4\mu} \left[ \epsilon^{\mu\nu\rho}\left(R_{\mu\nu}{}^{ab} \omega_{\rho a b} + \frac23 \omega_\mu{}^{ab} \omega_{\nu b}{}^c \omega_{\rho c a}\right) - 8 \epsilon^{\mu \nu \rho} V_\mu \partial_\nu V_\rho \right] \nonumber \\
& \ \ \ + \frac{1}{m^2} \Big[R_{\mu\nu}R^{\mu\nu} - \frac38 R^2 - R_{\mu\nu}V^\mu V^\nu - F_{\mu \nu}F^{\mu\nu} + \frac14 R(V^2 - B^2) \nonumber \\
& \ \ \ \ \ \  +\frac16 |S|^2(A^2 - 4 B^2) - \frac12 V^2 (3 A^2 + 4 B^2) - 2 V^\mu B \partial_\mu A \Big] \,,
\end{align}
where $F_{\mu\nu}$ denotes the field strength of $V_\mu$ and $A$ and $B$ are the real and imaginary parts of the auxiliary field $S$:
\begin{equation}
  \label{eq:defAB}
  S = A + \rmi B \,.
\end{equation}
The bosonic part of ordinary $\mathcal{N}=(1,1)$ supergravity, in the presence of a cosmological constant is obtained from (\ref{N11GMGnl}) by sending 
$(\mu$, $m^2) \rightarrow \infty$. Similarly, the limit $m^2 \rightarrow \infty$ leads to $\mathcal{N}=(1,1)$ Topologically Massive Supergravity and the limit 
$\mu \rightarrow \infty$ corresponds to the $\mathcal{N}=(1,1)$ supergravity version of 
New Massive Gravity (NMG)\cite{Bergshoeff:2009hq}. The bosonic part of (off-shell) $\mathcal{N}=(1,0)$ Topologically Massive Supergravity studied in 
\cite{Becker:2009mk} can be found from (\ref{N11GMGnl}) by sending $m^2 \rightarrow \infty$, truncating the vector field $V_\mu$, putting the scalar 
field $B=0$ and restricting all spinors to be Majorana instead of Dirac. 

This theory admits a maximally supersymmetric AdS$_3$ background given by
\begin{equation}
  \label{eq:vacuum}
  \bar{R}_{\mu\nu} = -\frac{2}{\ell^2} \bar{g}_{\mu\nu} \,, \qquad \bar{A} = -\frac{1}{\ell} \,,  \qquad \bar{B} = 0 \,, \qquad \bar{V}_\mu = 0 \,,
\end{equation}
where here and in the following, we will denote background quantities with a bar and $\ell$ is the AdS length. It is related to the cosmological 
constant $\Lambda$ via $\Lambda = -1/\ell^2$ and to the parameters appearing in (\ref{N11GMGnl}) via
\begin{equation}
  \label{eq:adslength}
  4 \sigma + \ell M + \frac{2}{3 \ell^2 m^2} = 0 \,.
\end{equation}

The fermionic terms of the $\mathcal{N}=(1,1)$ General Massive Supergravity were not given explicitly in \cite{Alkac:2014hwa}. Since, in this 
paper we will be concerned with studying solutions of all equations of motion, linearized around the background (\ref{eq:vacuum}), we are also interested 
in the linearization of the fermionic terms\footnote{For $\mathcal{N}=1$ TMG, these terms were worked out in \cite{Becker:2009mk}. For the $\mathcal{N}=(1,1)$ theory at hand, the linearization and mode spectrum were discussed using superfields and superspace in \cite{Kuzenko:2015jda}. The fermionic terms given here, as well as the analysis of fermionic modes are therefore implicit in \cite{Kuzenko:2015jda}.}. We will obtain these in the 
next section, by supersymmetrizing the linearization of the bosonic action (\ref{N11GMGnl}).

\subsection{Linearized $\mathcal{N}=(1,1)$ General Massive Supergravity}

In this section, we will construct linearized $\mathcal{N}=(1,1)$ General Massive Supergravity, starting from the linearization of the action (\ref{N11GMGnl}). We therefore split all bosonic fields in their background values (\ref{eq:vacuum}) and small fluctuations
\begin{alignat}{2}
  \label{eq:linansatz}
  g_{\mu\nu} &= \bar{g}_{\mu\nu} + \kappa h_{\mu \nu} \,, \qquad & V_\mu &= \kappa v_\mu \,, \nonumber \\
  S &= -\frac{1}{\ell} + \kappa s = -\frac{1}{\ell} + \kappa \left(a + \rmi b\right) \,,
\end{alignat}
where $\kappa$ is the gravitational coupling constant.
Using this ansatz, we find the following Lagrangian for the bosonic part of linearized $\mathcal{N}=(1,1)$ General Massive Supergravity
\begin{align} \label{eq:linLagrbos}
  \bar{e}^{-1}\mathcal{L}_{\mathrm{bos}} &= \left(-\frac{\sigma}{2} + \frac{1}{4 m^2 \ell^2}\right) h^{\mu\nu}\mathcal{G}_{\mu\nu}(h) - \frac{1}{2\mu}h^{\mu\nu} \mathcal{C}_{\mu\nu}(h) - \frac{1}{2 m^2} h^{\mu\nu}\mathcal{K}_{\mu\nu}(h) + \frac{2}{\mu} \epsilon^{\mu\nu\rho} v_\mu \partial_\nu v_\rho \nonumber \\ & \ \ \ + \left(\frac{1}{m^2 \ell^2} - 2\sigma \right) \left(a^2 + b^2\right) - \frac{1}{m^2} f_{\mu\nu}f^{\mu\nu} - \left(\frac{1}{m^2 \ell^2} - 2\sigma \right) v^2 \,.
\end{align}
In this Lagrangian, we have introduced the linearized Einstein tensor $\mathcal{G}_{\mu\nu}(h)$, Cotton tensor $\mathcal{C}_{\mu\nu}(h)$ and a tensor $\mathcal{K}_{\mu\nu}(h)$. The first two are defined as
\begin{align} \label{eq:linEinstCotton} 
\mathcal{G}_{\mu\nu}(h) &=
R^{(1)}_{\mu\nu} - \frac12 \bar{g}_{\mu\nu} \bar{g}^{\rho\sigma}
R^{(1)}_{\rho\sigma} - 2 \Lambda h_{\mu\nu} + \Lambda \bar{g}_{\mu\nu}
h\,, \nonumber \\
\mathcal{C}_{\mu\nu}(h) &= \epsilon_{\mu}{}^{\tau\rho}\bar{\nabla}_{\tau}\left(R^{(1)}_{\rho\nu}-\frac14 \bar{g}_{\rho\nu}\bar{g}^{\alpha\beta}R^{(1)}_{\alpha\beta}-2\Lambda h_{\rho\nu}+\frac{\Lambda}{2} \bar{g}_{\rho\nu}h\right) \,,  
\end{align}
where
\begin{align}
  R^{(1)}_{\mu\nu} &= -\frac12\left(\bar{\nabla}^\rho \bar{\nabla}_\rho
    h_{\mu\nu}-\bar{\nabla}^\rho\bar{\nabla}_\mu h_{\rho\nu} -
    \bar{\nabla}^\rho \bar{\nabla}_\nu h_{\rho\mu} + \bar{\nabla}_\mu
    \bar{\nabla}_\nu h\right) \,, 
\end{align} 
$h=\bar{g}^{\mu\nu} h_{\mu\nu}$ and $\bar{\nabla}_\mu$ denotes a derivative that is covariantized with respect to the background Levi-Civita connection.
The tensor $\mathcal{K}_{\mu\nu}(h)$ is then defined in terms of the Cotton tensor via
\begin{equation} \label{eq:linKtens}
  \mathcal{K}_{\mu\nu}(h) = \epsilon_{\mu}{}^{\tau\rho}\bar{\nabla}_\tau \mathcal{C}_{\rho\nu}(h)\,.
\end{equation}
The field strength $f_{\mu\nu}$ of $v_\mu$ is defined in the usual way
\begin{equation}
  \label{eq:deffmunu}
  f_{\mu\nu} = \bar{\nabla}_{\mu} v_\nu - \bar{\nabla}_{\nu} v_\mu \,.
\end{equation}
The linearized action (\ref{eq:linLagrbos}) is invariant under the following linearized diffeomorphisms
\begin{equation}
  \label{eq:lindiff}
  \delta h_{\mu\nu} = \bar{\nabla}_\mu \xi_\nu + \bar{\nabla}_\nu \xi_\mu \,.
\end{equation}
In particular, $\mathcal{G}_{\mu\nu}(h)$, $\mathcal{C}_{\mu\nu}(h)$ and $\mathcal{K}_{\mu\nu}(h)$ are invariant under this gauge transformation.

Using the ansatz (\ref{eq:linansatz}), supplemented with
\begin{equation}
  \label{eq:lingravitino}
  \Psi_\mu = \kappa \psi_\mu \,,
\end{equation}
one can linearize the transformation rules (\ref{eq:susyrulesnl}). The result is 
\begin{align} \label{eq:linsusys}
\delta h_{\mu\nu} &= \bar{\epsilon} \gamma_{(\mu} \psi_{\nu)} + \mathrm{h.c.} \,, \nonumber \\
\delta v_\mu &= \frac{\rmi}{8} \bar{\epsilon} \gamma^{\nu\rho}\gamma_\mu \psi_{\nu\rho} + \frac{\rmi}{4\ell} \bar{\epsilon} (B \psi_\mu)^* + \mathrm{h.c.} \,, \nonumber \\
\delta s &= -\frac14 \tilde{\epsilon} \gamma^{\mu\nu}\psi_{\mu\nu} + \frac{1}{2\ell}\tilde{\epsilon} \gamma^\mu (B\psi_\mu)^*\,, \nonumber \\
\delta \psi_\mu &= -\frac14 \gamma^{\rho\sigma} \bar{\nabla}_\rho h_{\mu\sigma} \epsilon - \frac{\rmi}{2} v_\nu \gamma^\nu \gamma_\mu \epsilon + \frac{1}{4\ell} h_{\mu\nu}\gamma^\nu (B\epsilon)^* - \frac12 s \gamma_\mu (B\epsilon)^* \,,
\end{align}
where $\psi_{\mu\nu} = 2 \bar{D}_{[\mu} \psi_{\nu]}$ (with $\bar{D}_\mu$ the spinor derivative that is covariantized with respect to the background spin connection) and the supersymmetry parameter $\epsilon$ satisfies the Killing spinor equation:
\begin{equation}
  \label{eq:killspinor}
  \bar{D}_\mu \epsilon + \frac{1}{2\ell} \gamma_\mu (B\epsilon)^* = 0 \,, \qquad \bar{D}_\mu (B\epsilon)^* + \frac{1}{2\ell} \gamma_\mu \epsilon = 0 \,.
\end{equation}
In order to find the supersymmetric completion of the linearized action (\ref{eq:linLagrbos}), we define the following tensors
\begin{align} \label{eq:fermtensors}
\tilde{\mathcal{R}}_\mu &\equiv \mathcal{R}_\mu - \frac{1}{2\ell} \gamma_\mu{}^\nu (B \psi_\nu)^* \equiv \epsilon_\mu{}^{\nu\rho} \bar{D}_\nu \psi_\rho  - \frac{1}{2\ell} \gamma_\mu{}^\nu (B \psi_\nu)^*\,, \nonumber \\
\mathcal{C}_\mu &\equiv \epsilon_\mu{}^{\nu\rho} \bar{D}_\nu \mathcal{R}_\rho + \gamma^\nu \bar{D}_\nu \mathcal{R}_\mu - \frac{1}{2\ell^2} \psi_\mu \,, \nonumber \\
\mathcal{K}_\mu &\equiv \gamma^\nu \bar{D}_\nu \mathcal{C}_\mu \,.
\end{align}
These tensors are the fermionic equivalents of the bosonic tensors (\ref{eq:linEinstCotton}) and
(\ref{eq:linKtens}) that are defined in terms of the metric perturbation. They are invariant under the
following fermionic symmetry:
\begin{equation}
  \label{eq:zetasymm}
  \delta \psi_\mu = \bar{D}_\mu \zeta + \frac{1}{2\ell} \gamma_\mu (B\zeta)^* \,,
\end{equation}
that is a remnant of local supersymmetry transformations. One can use them to construct the fermionic terms in the supersymmetric 
completion of (\ref{eq:linLagrbos}). In order to do so, one needs their variations under the linearized supersymmetries (\ref{eq:linsusys}). We find:
\begin{align}
\delta \tilde{\mathcal{R}}_\mu &= \frac12 \gamma^\nu \mathcal{G}_{\nu\mu}(h) \epsilon - \frac{\rmi}{2} \gamma^\sigma \gamma_\mu{}^\rho \bar{\nabla}_\rho v_\sigma \epsilon + \frac{\rmi}{\ell} v_\mu (B \epsilon)^* - \frac{\rmi}{2\ell} \gamma_\mu{}^\nu v_\nu (B\epsilon)^*\nonumber \\ & \ \ \  - \frac12 \gamma_\mu{}^\nu \partial_\nu s (B \epsilon)^* + \frac{1}{2\ell} (s + \bar{s}) \gamma_\mu \epsilon \,, \nonumber \\
\delta \mathcal{C}_\mu &= \gamma^\nu \mathcal{C}_{\nu \mu}(h)\epsilon - \rmi \bar{\nabla}^\nu f_{\nu\mu} \epsilon - \frac{\rmi}{2} \epsilon_\mu{}^{\nu\rho} \gamma^\sigma \bar{\nabla}_\sigma f_{\nu\rho}\epsilon + \frac{\rmi}{\ell} \epsilon_\mu{}^{\nu\rho} f_{\nu\rho}(B\epsilon)^* \nonumber \\
& \ \ \ + \frac{\rmi}{\ell} \gamma^\nu f_{\nu\mu} (B\epsilon)^* \,, \nonumber \\
\delta \mathcal{K}_\mu &= \gamma^\nu \mathcal{K}_{\nu\mu}(h) \epsilon + \frac{1}{2\ell} \gamma^\nu \mathcal{C}_{\nu\mu}(h) (B\epsilon)^* - \rmi \gamma^\nu \bar{\nabla}_\nu \bar{\nabla}^\rho f_{\rho\mu} \epsilon - \frac{\rmi}{2} \epsilon_\mu{}^{\alpha\beta} \bar{\nabla}^\rho \bar{\nabla}_\rho f_{\alpha\beta}\epsilon \nonumber \\ & \ \ \ + \frac{5\rmi}{2\ell} \bar{\nabla}^\nu f_{\nu\mu} (B\epsilon)^* + \frac{3\rmi}{4\ell} \epsilon_\mu{}^{\alpha\beta} \gamma^\rho \bar{\nabla}_\rho f_{\alpha\beta} (B\epsilon)^* + \frac{\rmi}{\ell} \epsilon_\rho{}^{\alpha\beta} \gamma^\rho \bar{\nabla}_\alpha f_{\beta\mu} (B\epsilon)^* \nonumber \\ & \ \ \ -\frac{3\rmi}{2\ell^2} \epsilon_\mu{}^{\alpha\beta}f_{\alpha\beta}\epsilon - \frac{\rmi}{2\ell^2} \gamma^\nu f_{\nu\mu} \epsilon\,.
\end{align}
Using the above transformation rules, it can be checked that the following three expressions are supersymmetric invariants:
\begin{align} \label{susyinvariants}
\bar{e}^{-1} \mathcal{L}_{\mathrm{Einst}} &= -\frac{\sigma}{2} h^{\mu\nu} \mathcal{G}_{\mu\nu}(h) - 2 \sigma |s|^2 + 2 \sigma v_\mu v^\mu - \sigma \left( \bar{\psi}^\mu \Big( \mathcal{R}_\mu - \frac{1}{2\ell} \gamma_\mu{}^\nu (B\psi_\nu)^* \Big) + \mathrm{h.c.} \right) \,, \nonumber \\ \bar{e}^{-1} \mathcal{L}_{\mathrm{TMG}} &= -\frac{1}{2\mu} h^{\mu\nu} \mathcal{C}_{\mu\nu}(h) + \frac{2}{\mu} \epsilon^{\mu\nu\rho} v_\mu \partial_\nu v_\rho - \frac{1}{2\mu} \left( \bar{\psi}^\mu \mathcal{C}_\mu + \mathrm{h.c.}\right) \,, \nonumber \\  \bar{e}^{-1} \mathcal{L}_{\mathrm{NMG}} &= -\frac{1}{2 m^2} h^{\mu\nu} \mathcal{K}_{\mu\nu}(h) - \frac{1}{m^2} f_{\mu \nu} f^{\mu\nu} - \frac{1}{2 m^2} \left( \bar{\psi}^\mu \Big( \mathcal{K}_\mu - \frac{1}{2\ell} (B \mathcal{C}_\mu)^* \Big) \Big) + \mathrm{h.c.} \right) \,,
\end{align}
and the supersymmetrization of \eqref{eq:linLagrbos} is therefore 
\begin{align}
  \label{eq:linLagrsusy}
  \bar{e}^{-1} \mathcal{L} &= \bar{e}^{-1} \left( \Big(1 - \frac{1}{2 m^2 \ell^2 \sigma} \Big) \mathcal{L}_{\mathrm{Einst}} + \mathcal{L}_{\mathrm{TMG}} + \mathcal{L}_{\mathrm{NMG}} \right) \,, \nonumber \\
&= \left(-\frac{\sigma}{2} + \frac{1}{4 m^2 \ell^2}\right) h^{\mu\nu} \mathcal{G}_{\mu\nu}(h) - \frac{1}{2\mu} h^{\mu\nu} \mathcal{C}_{\mu\nu}(h) - \frac{1}{2 m^2} h^{\mu\nu} \mathcal{K}_{\mu\nu}(h) \nonumber \\ & \ \ \ - \left(2\sigma - \frac{1}{m^2 \ell^2} \right) |s|^2 + \left( 2\sigma - \frac{1}{m^2 \ell^2} \right) v^2 + \frac{2}{\mu} \epsilon^{\mu\nu\rho} v_\mu \partial_\nu v_\rho - \frac{1}{m^2} f_{\mu\nu} f^{\mu \nu} \nonumber \\ & \ \ \ + \bigg[ \left(-\sigma + \frac{1}{2 m^2 \ell^2} \right) \bar{\psi}^\mu \left(\mathcal{R}_\mu - \frac{1}{2\ell} \gamma_\mu{}^\nu (B \psi_\nu)^* \right) - \frac{1}{2\mu} \bar{\psi}^\mu \mathcal{C}_\mu \nonumber \\ & \ \ \ - \frac{1}{2m^2} \bar{\psi}^\mu \left( \mathcal{K}_\mu - \frac{1}{2\ell} (B \mathcal{C}_\mu)^* \right) + \mathrm{h.c.} \bigg]
\end{align}

\section{The non-critical spectrum} \label{sec:noncritspectrum}

In this section, we will study the spectrum of linearized modes propagated by the Lagrangian (\ref{eq:linLagrsusy}), for generic values of its 
parameters. The equations of motion derived from (\ref{eq:linLagrsusy}) are:
\begin{align}
  & \frac{\Omega}{\ell^2 m^2} \mathcal{G}_{\mu\nu}(h) + \frac{1}{\mu} \mathcal{C}_{\mu\nu}(h) + \frac{1}{m^2} \mathcal{K}_{\mu\nu}(h) = 0 \,, \label{eq:heq} \\
& \frac{\Omega}{\ell^2 m^2} s = 0 \,, \label{eq:seq} \\
& \frac{\Omega}{\ell^2 m^2} v_\mu + \frac{1}{\mu} \epsilon_\mu{}^{\nu\rho} \partial_\nu v_\rho + \frac{1}{m^2} \bar{\nabla}^\nu f_{\nu\mu} = 0 \,, \label{eq:veq} \\
& \frac{\Omega}{\ell^2 m^2} \tilde{\mathcal{R}}_\mu + \frac{1}{2\mu} \mathcal{C}_\mu + \frac{1}{2m^2} \left(\mathcal{K}_\mu - \frac{1}{2\ell} (B \mathcal{C}_\mu)^* \right) = 0 \,, \label{eq:psieq}
\end{align}
where we have defined
\begin{equation}
  \label{eq:defOmega}
  \Omega = \sigma \ell^2 m^2 - \frac12 \,.
\end{equation}
For $\Omega = 0$, the scalar field $s$ does not appear in the Lagrangian (\ref{eq:linLagrsusy}), while for $\Omega \neq 0$ one finds that $s = 0$. In either 
case, there are no propagating modes associated to $s$. The modes propagated by the other equations are best analyzed by adopting certain gauge choices that 
fix the residual linearized diffeomorphisms (\ref{eq:lindiff}) and local fermionic symmetry (\ref{eq:zetasymm}). We will now discuss each of these in turn.

\subsection{Graviton spectrum}

In order to discuss the physical modes described by (\ref{eq:heq})--(\ref{eq:psieq}), we will choose the transverse-traceless
gauge that fixes the linearized diffeomorphisms (\ref{eq:lindiff}):
\begin{equation}
  \label{eq:ttgauge}
  \bar{\nabla}^\nu h_{\mu\nu} = 0 \,, \qquad h = 0 \,.
\end{equation}
This gauge choice is always possible. In particular, one can always consider a family of gauges, parametrized by a constant $c$ \cite{Lu:2011mw}:
\begin{equation}
  \label{eq:diffchoicec}
  \bar{\nabla}^\nu h_{\mu \nu} = c\, \bar{\nabla}_\mu h \,.
\end{equation}
Using this gauge choice in the trace of (\ref{eq:heq}), one gets
\begin{equation} \label{eq:tracecheq}
  (1-c) \bar{\nabla}^\mu \bar{\nabla}_\mu h - \frac{2}{\ell^2} h = 0 \,.
\end{equation}
For $c=1$, one sees that (\ref{eq:diffchoicec}) reduces to the transverse-traceless gauge as a consequence of the equation of motion. For $c \neq 1$, one can easily see that linearized diffeomorphisms with parameter $\xi_\mu = \partial_\mu \xi$, where $\xi$ obeys
\begin{equation}
  (1-c) \bar{\nabla}^\mu \bar{\nabla}_\mu \xi - \frac{2}{\ell^2} \xi = 0 \,,
\end{equation}
preserve the gauge choice (\ref{eq:diffchoicec}). This residual gauge transformation acts on $h$ as $\delta h = 2 \bar{\nabla}^\mu \bar{\nabla}_\mu \xi \sim \xi$ and can thus be fixed by setting $h = 0$, showing that one can again adopt a transverse-traceless gauge.

Adopting the gauge choice (\ref{eq:ttgauge}), the equation of motion (\ref{eq:heq}) simplifies and can be most easily written down using the differential operators
\begin{equation}
  \mathcal{D}(\eta)_\mu{}^\nu = \frac{1}{\ell} \delta_\mu^\nu + \frac{\eta}{\sqrt{|\bar{g}|}} \epsilon_\mu{}^{\tau\nu}\bar{\nabla}_\tau \,.
\end{equation}
In terms of these, the gauge fixed field equation for $h_{\mu\nu}$ can be written as
\begin{align} \label{eq:fixedmetriceq}
&  \left(\mathcal{D}(1)\mathcal{D}(-1)\mathcal{D}(\eta_1)\mathcal{D}(\eta_2)h\right)_{\mu\nu}=0\,,
\end{align}
where $\eta_{1,2}$ obey
\begin{equation}
  \eta_1\eta_2 = \frac{1}{\Omega} \,, \qquad \eta_1 + \eta_2 = \frac{\ell m^2}{\mu \Omega}\,.
\end{equation}
For generic parameter values, $\eta_1$ and $\eta_2$ will be different from each other and from $\pm 1$. If this is the case, we see from (\ref{eq:fixedmetriceq}) that the solution spectrum generically contains 2 `massless' graviton modes $h^L_{\mu\nu}$, $h^R_{\mu\nu}$ obeying
\begin{align}
  \left(\mathcal{D}(1) h^L\right)_{\mu\nu} = 0 \,, \qquad \left(\mathcal{D}(-1) h^R\right)_{\mu\nu} = 0 \,.
\end{align}
These modes are non-propagating in the AdS$_3$ bulk, but can lead to interesting dynamics on the boundary of AdS$_3$. For this reason, they are often called `boundary gravitons'. In addition to these, the equation (\ref{eq:fixedmetriceq}) also propagates two massive graviton modes $h_{\mu\nu}(\eta_{1,2})$ that obey
\begin{align}
  \left(\mathcal{D}(\eta_{1,2}) h(\eta_{1,2})\right)_{\mu\nu} = 0  \,.
\end{align}
We will discuss what happens for specific parameter values for which $\{1, -1$, $\eta_1$, $\eta_2\}$ are not all different in section \ref{sec:critpoints}. 

\subsection{Gravitino spectrum} \label{ssec:gravitinospec}

In order to discuss the gravitino spectrum, we have to fix the fermionic symmetry (\ref{eq:zetasymm}). This can be done by adopting the gauge choice
\begin{equation}
  \label{eq:gtrless}
  \gamma^\mu \psi_\mu = 0 \,.
\end{equation}
Contracting the gravitino equation of motion (\ref{eq:psieq}) with $\gamma^\mu$, one finds
\begin{equation}
  \left(-\sigma + \frac{3}{2 m^2 \ell^2} \right) \slashed{\bar{D}} (\gamma^\mu \psi_\mu) + \frac{1}{\ell} \left( \sigma - \frac{1}{2 m^2 \ell^2} \right) (B \gamma^\mu \psi_\mu)^* - \left(- \sigma + \frac{3}{2 m^2 \ell^2} \right) \bar{D}^\mu \psi_\mu = 0 \,,
\end{equation}
so that, generically, the gauge choice (\ref{eq:gtrless}) together with the gravitino equation of motion implies that 
\begin{equation}
  \label{eq:gravdivless}
  \bar{D}^\mu \psi_\mu = 0 \,.
\end{equation}
Upon using (\ref{eq:gtrless}) and (\ref{eq:gravdivless}), the gravitino equation of motion simplifies to
\begin{align}
  \label{eq:fixedgraveq}
  & -\frac{\Omega}{\ell^2 m^2} \left(\slashed{\bar{D}} \psi_\mu + \frac{1}{2\ell} (B \psi_\mu)^* \right) - \frac{1}{\mu} \left(\slashed{\bar{D}} \slashed{\bar{D}} - \frac{1}{4 \ell^2}\right) \psi_\mu - \frac{1}{m^2} \slashed{\bar{D}} \left(\slashed{\bar{D}} \slashed{\bar{D}} - \frac{1}{4\ell^2} \right) \psi_\mu \nonumber \\ & \qquad + \frac{1}{2 m^2 \ell} \left(\slashed{\bar{D}} \slashed{\bar{D}} - \frac{1}{4\ell^2} \right) (B \psi_\mu)^* = 0 \,.
\end{align}
It is useful to split the Dirac gravitino $\psi_\mu$ in two Majorana gravitini $\psi_{\mu\, 1,2}$, with $\psi_\mu = \psi_{\mu\, 1} + \rmi \psi_{\mu\, 2}$ as in (\ref{eq:diracsplit}). The gravitino equation of motion (\ref{eq:fixedgraveq}) can then be factorized as
\begin{align}
  \label{eq:fixedgraveqfact}
  &  \left( \slashed{\bar{D}} + \frac{1}{2\ell} \right) \left( \slashed{\bar{D}} - \frac{1}{2 \ell} + \frac{1}{\eta_1 \ell} \right) \left( \slashed{\bar{D}} - \frac{1}{2\ell} + \frac{1}{\eta_2 \ell} \right) \psi_{\mu\, 1} = 0 \,, \nonumber \\
  & \left( \slashed{\bar{D}} - \frac{1}{2\ell} \right) \left( \slashed{\bar{D}} + \frac{1}{2 \ell} + \frac{1}{\eta_1 \ell}  \right) \left( \slashed{\bar{D}} + \frac{1}{2\ell} + \frac{1}{\eta_2 \ell} \right) \psi_{\mu\, 2} = 0 \,.
\end{align}
Generically, the spectrum thus contains 2 `massless' boundary gravitino modes $\psi^L_\mu$, $\psi^R_\mu$ that obey
\begin{align}
  \label{eq:nomassgrav}
  \left(\slashed{\bar{D}} + \frac{1}{2\ell}\right) \psi_{\mu}^L = 0 \,, \qquad \left(\slashed{\bar{D}} - \frac{1}{2\ell}\right) \psi_{\mu}^R = 0 \,.
\end{align}
Similar to the graviton case, these modes are non-propagating in the AdS$_3$ bulk but can nevertheless have non-trivial boundary dynamics. In addition to these, one finds four propagating massive Majorana modes $\psi_{\mu}^-(\eta_{1,2})$, $\psi_{\mu}^+(\eta_{1,2})$ that are solutions of the equations
\begin{align}
  \label{eq:massgrav1}
  \left( \slashed{\bar{D}} - \frac{1}{2\ell} + \frac{1}{\eta_{1,2} \ell} \right) \psi_{\mu}^-(\eta_{1,2}) = 0 \,, \qquad \left( \slashed{\bar{D}} + \frac{1}{2\ell} + \frac{1}{\eta_{1,2} \ell} \right) \psi_{\mu}^+(\eta_{1,2}) = 0 \,.
\end{align}
Note that the modes $\psi_\mu^L$ and $\psi_\mu^-(\eta_{1,2})$ are solutions for the real part $\psi_{\mu\, 1}$ of the Dirac gravitino $\psi_\mu$, while the 
modes $\psi_\mu^R$ and $\psi_\mu^+(\eta_{1,2})$ are solutions for the imaginary part $\psi_{\mu\, 2}$. The massless mode $\psi_{\mu}^L$ can be obtained from the 
massive modes $\psi_\mu^-(\eta_{1,2})$ by putting $\eta_{1,2} = 1$: $\psi_{\mu}^-(1) = \psi_\mu^L$. Similarly, the massless mode $\psi_{\mu}^R$ is obtained from the 
massive modes $\psi_\mu^+(\eta_{1,2})$ by putting $\eta_{1,2} = -1$: $\psi_{\mu}^+(-1) = \psi_\mu^R$. For generic parameter values however, all the above gravitino modes are distinct. We will discuss what happens at special parameter values for which some of these modes seemingly coincide in section 4.

\subsection{Vector spectrum}

The equation of motion (\ref{eq:veq}) for $v_\mu$ can be simplified for generic parameter values. For $\Omega \neq 0$, contracting (\ref{eq:veq}) with $\bar{\nabla}^\mu$ leads to
\begin{align}
  \label{eq:consveceq}
  \frac{\Omega}{\ell^2 m^2} \bar{\nabla}^\mu v_\mu = 0 \,,
\end{align}
so that $v_\mu$ is divergence-less
\begin{align}
  \label{eq:consveceqshort}
  \bar{\nabla}^\mu v_\mu = 0 \,,
\end{align}
as a consequence of its equation of motion.
If $\Omega = 0$, the equation (\ref{eq:veq}) becomes invariant under an accidental U(1) gauge symmetry $\delta v_\mu = \partial_\mu \Lambda$ and one can still impose (\ref{eq:consveceqshort}) as a gauge fixing condition.
Using this, one finds that (\ref{eq:veq}) simplifies to 
\begin{align}
  \label{eq:simpleveceq}
  \left( \frac{\Omega + 2}{\ell^2 m^2} \right) v_\mu + \frac{1}{\mu} \epsilon_\mu{}^{\nu\rho} \partial_\nu v_\rho + \frac{1}{m^2} \bar{\nabla}^\nu \bar{\nabla}_\nu v_\mu = 0 \,.
\end{align}
This can be rewritten as
\begin{align}
  \label{eq:simpleveceq2}
  \frac{1}{m^2 \eta_1 \eta_2} \left(\mathcal{D}(\eta_1) \mathcal{D}(\eta_2) v\right)_\mu = 0 \,.
\end{align}
Summarizing, we find that the original vector equation of motion generically is equivalent to
\begin{align}
  \label{eq:equivveceqs}
  \left(\mathcal{D}(\eta_1) \mathcal{D}(\eta_2) v \right)_\mu = 0 \,, \qquad \bar{\nabla}^\mu v_\mu = 0 \,.
\end{align}
One can thus see that the physical spectrum contains 2 massive vector modes $v_\mu(\eta_{1,2})$, that obey
\begin{align}
  \left(\mathcal{D}(\eta_{1,2})v(\eta_{1,2})\right)_\mu = 0 \,.
\end{align}
Note that there are no massless vector modes, unlike for the graviton and gravitino.

\subsection{Supersymmetry properties}

The linearized modes of the previous subsection naturally organize themselves in two massless supermultiplets $\{h_{\mu\nu}^{L/R}, \psi_\mu^{L/R}\}$ and two massive supermultiplets $\{h_{\mu\nu}(\eta_{1,2})$, $\psi_\mu^-(\eta_{1,2})$, $\psi_\mu^+(\eta_{1,2})$, $v_\mu(\eta_{1,2})\}$. We will now calculate the supersymmetry transformation rules that connect the various modes in each supermultiplet. Note that the gauge fixing conditions that were imposed in the previous subsections are generically not left invariant under the linearized supersymmetry transformations of (\ref{eq:linsusys}). These transformation rules therefore have to be modified with compensating linearized diffeomorphisms (\ref{eq:lindiff}) and fermionic symmetries (\ref{eq:zetasymm}), such that the modified transformations preserve all gauge choices. In this subsection, we will first obtain the required compensating transformations. We will then show how the linearized modes transform into each other under the modified supersymmetry transformations.

\subsubsection{Compensating transformations} \label{sssec:compensatingtrafos}

Since the analysis of physical modes of section \ref{ssec:gravitinospec} was done using Majorana spinors instead of Dirac spinors, we will similarly study their supersymmetry properties in terms of real Majorana spinors. Splitting the Dirac spinors into real and imaginary parts as in eq. (\ref{eq:diracsplit}), one finds that the supersymmetry transformation of the metric perturbation given in (\ref{eq:linsusys}) can be rewritten as:
\begin{equation}
  \label{eq:deltahreal}
  \delta h_{\mu\nu} = 2 \bar{\epsilon}_1 \gamma_{(\mu} \psi_{\nu)1} + 2 \bar{\epsilon}_2 \gamma_{(\mu} \psi_{\nu)2} \,.
\end{equation}
One then immediately sees that the gauge condition $h=0$ is invariant under supersymmetry, by virtue of the gauge conditions $\gamma^\mu\psi_{\mu\, 1,2} = 0$. 
The gauge condition $\bar{\nabla}^\nu h_{\mu\nu}=0$ is however not preserved by supersymmetry. In order to maintain its invariance under supersymmetry, the 
rule (\ref{eq:deltahreal}) thus needs to be modified with a compensating diffeomorphism. The transformation rules of the massless multiplets can be obtained as a 
special case of those of the massive ones. In order to calculate the required compensating transformations, we will thus first focus on the massive multiplets. In that 
case, $h_{\mu\nu}$ corresponds to the modes $h_{\mu\nu}(\eta_{1,2})$ and the gravitini $\psi_{\mu\, 1}$, $\psi_{\mu\, 2}$ correspond to modes 
$\psi_{\mu}^-(\eta_{1,2})$, $\psi_{\mu}^+(\eta_{1,2})$ 
respectively. The following does not depend on which of the $\eta_{1,2}$ is considered, so we will for simplicity denote $\eta_{1,2}$ by $\eta$ below. 
One finds that under supersymmetry
\begin{align} \label{eq:deltadivh}
 \delta\left( \bar{\nabla}^\nu h_{\mu\nu}(\eta)\right) &= \frac{1}{\ell} \left(3 - \frac{1}{\eta} \right) \bar{\epsilon}_1 \psi_{\mu}^-(\eta) - \frac{1}{\ell} \left(3 + \frac{1}{\eta} \right) \bar{\epsilon}_2 \psi_{\mu}^+(\eta) \nonumber \\ & \quad + \bar{\epsilon}_1 \left(\slashed{\bar{D}} - \frac{1}{2\ell}  + \frac{1}{\eta \ell}\right) \psi_{\mu}^-(\eta) + \bar{\epsilon}_2 \left(\slashed{\bar{D}} + \frac{1}{2\ell} + \frac{1}{\eta \ell}\right) \psi_{\mu}^+(\eta) \,. 
\end{align}
This transformation rule is obtained without using the linearized equations of motion. Note that the last two terms of (\ref{eq:deltadivh}) are zero on-shell, 
according to equation (\ref{eq:massgrav1}). In the next section we will be interested in critical points in parameter space, where the gravitini no longer obey a 
first order equation and these terms can then no longer be assumed to be zero. As we will explain later, the supersymmetry transformation rules of the modes at these critical 
points can be obtained from those away from the critical points via a limiting procedure. This procedure requires that one keeps track of terms (up to a certain 
order in derivatives) in the compensating 
transformations that are zero on-shell away from the critical points. In order to simplify the discussion later, we will in this section already keep track of such 
terms. One can then propose the following ansatz for the parameter $\xi_\mu$ of compensating diffeomorphisms
\begin{align}
  \label{eq:diffcompansatz}
  \xi_\mu(\eta) &= a_1 \bar{\epsilon}_1 \psi_{\mu}^-(\eta) + b_1 \bar{\epsilon}_1 \left(\slashed{\bar{D}} - \frac{1}{2\ell} + \frac{1}{\eta\ell}\right) \psi_{\mu}^-(\eta) + c_1 \bar{\epsilon}_1 \left( \slashed{\bar{D}} - \frac{1}{2\ell} + \frac{1}{\eta\ell}\right)^2 \psi_{\mu}^-(\eta) \nonumber \\ & \quad + a_2 \bar{\epsilon}_2 \psi_{\mu}^+(\eta) + b_2 \bar{\epsilon}_2 \left(\slashed{\bar{D}} + \frac{1}{2\ell} + \frac{1}{\eta\ell}\right) \psi_{\mu}^+(\eta) + c_2 \bar{\epsilon}_2 \left( \slashed{\bar{D}} + \frac{1}{2\ell} + \frac{1}{\eta\ell}\right)^2 \psi_{\mu}^+(\eta)\,,
\end{align}
where we have included terms that are zero on-shell, for the reason explained above. As will be outlined in section \ref{sec:critpoints}, for the calculation of the transformation rules at the critical points, it will be sufficient to keep such terms up to second order in derivatives and assume that 
\begin{align} \label{eq:threeDpsizero}
\left( \slashed{\bar{D}} - \frac{1}{2\ell} + \frac{1}{\eta\ell}\right)^3 \psi_{\mu}^-(\eta) = 0 \,, \qquad  \left( \slashed{\bar{D}} + \frac{1}{2\ell} + \frac{1}{\eta\ell}\right)^3 \psi_{\mu}^+(\eta) = 0 \,.
\end{align}
The coefficients $a_1, \cdots, c_2$ appearing in (\ref{eq:diffcompansatz}) depend on $\eta$ and can be found by requiring that
\begin{align}
  \delta \left(\bar{\nabla}^\nu h_{\mu\nu}(\eta)\right) + 2 \bar{\nabla}^\nu \left(\bar{\nabla}_{(\mu} \xi_{\nu)}(\eta) \right) = 0\,,
\end{align}
upon using (\ref{eq:threeDpsizero}). The resulting diffeomorphism parameter is given by
\begin{align}
  \label{eq:compdiffeoresult}
 &\xi_{\mu}(\eta) = f(\eta) \bar{\epsilon}_1 \psi_\mu^-(\eta) + f(\eta)^2 \bar{\epsilon}_1 \left(\slashed{\bar{D}} - 
 \frac{1}{2\ell} + \frac{1}{\eta \ell} \right) \psi_\mu^-(\eta) + f(\eta)^3 \bar{\epsilon}_1 \left(\slashed{\bar{D}} - \frac{1}{2\ell} + 
 \frac{1}{\eta \ell} \right)^2 \psi_\mu^-(\eta) \nonumber \\ 
 &   - g(\eta) \bar{\epsilon}_2 \psi_\mu^+(\eta) + 
 g(\eta)^2 \bar{\epsilon}_2 \left(\slashed{\bar{D}} + \frac{1}{2\ell} + \frac{1}{\eta\ell} \right) \psi_\mu^+(\eta) - 
 g(\eta)^3 \bar{\epsilon}_2 \left(\slashed{\bar{D}} + \frac{1}{2\ell} + \frac{1}{\eta\ell} \right)^2 \psi_\mu^+(\eta)
\end{align}
where we have introduced the notation
\begin{align}
  f(\eta) \equiv \frac{\ell \eta}{\eta + 1} \,, \qquad g(\eta) \equiv \frac{\ell \eta}{\eta-1} \,.
\label{fg}
  \end{align}
One can discuss compensating transformations for the gravitini in a similar way. The transformation rule for the real and imaginary parts of the Dirac gravitino $\psi_\mu$ is given in terms of Majorana spinors by
\begin{align}
  \label{eq:Majgravrule}
  \delta \psi_{\mu\, 1} &= - \frac14 \gamma^{\rho\sigma} \bar{\nabla}_\rho h_{\mu \sigma} \epsilon_1 + \frac12 v_\nu \gamma^\nu \gamma_\mu \epsilon_2 + \frac{1}{4\ell} h_{\mu\nu} \gamma^\nu \epsilon_1 \,, \nonumber \\
 \delta \psi_{\mu\, 2} &= - \frac14 \gamma^{\rho\sigma} \bar{\nabla}_\rho h_{\mu \sigma} \epsilon_2 - \frac12 v_\nu \gamma^\nu \gamma_\mu \epsilon_1 - \frac{1}{4\ell} h_{\mu\nu} \gamma^\nu \epsilon_2 \,,
\end{align}
while the fermionic symmetry (\ref{eq:zetasymm}) is given by
\begin{align}
  \label{eq:zetasymmreal}
  \delta \psi_{\mu\, 1} &= \bar{D}_\mu \zeta_1 + \frac{1}{2\ell} \gamma_\mu \zeta_1 \,, \qquad \delta \psi_{\mu\, 2} = \bar{D}_\mu \zeta_2 - \frac{1}{2\ell} \gamma_\mu \zeta_2\,.
\end{align}
We will again focus on the massive multiplets first, for which $h_{\mu\nu}$ is given by $h_{\mu\nu}(\eta)$, $\psi_{\mu\, 1}$, $\psi_{\mu\, 2}$ correspond to $\psi_{\mu}^-(\eta)$, $\psi_\mu^+(\eta)$ and $v_\mu$ to $v_\mu(\eta)$. One finds that
\begin{align}
  \label{eq:vargaugepsiminus}
  \delta \left(\gamma^\mu \psi_\mu^-(\eta) \right) &= - \frac12 v_\mu(\eta) \gamma^\mu \epsilon_2 \,, \qquad  \qquad \delta \left(\gamma^\mu \psi_\mu^+(\eta) \right) =  \frac12 v_\mu(\eta) \gamma^\mu \epsilon_1 \,, \nonumber \\
  \delta \left(\bar{D}^\mu \psi_\mu^-(\eta) \right) &= \frac{1}{2\ell} \left(\frac{1}{\eta} + \frac32\right) v_\mu(\eta) \gamma^\mu \epsilon_2 - \frac{1}{2\eta} \gamma^\mu \left(\mathcal{D}(\eta) v(\eta)\right)_\mu \epsilon_2 \,, \nonumber \\
  \delta \left(\bar{D}^\mu \psi_\mu^+(\eta) \right) &= \frac{1}{2\ell} \left(\frac32 - \frac{1}{\eta}\right) v_\mu(\eta) \gamma^\mu \epsilon_1  + \frac{1}{2\eta} \gamma^\mu \left(\mathcal{D}(\eta) v(\eta)\right)_\mu \epsilon_1 \,.
\end{align}
Note that away from critical points the last terms in $\delta \left(\bar{D}^\mu \psi_\mu^\pm(\eta) \right)$ are zero on-shell but we keep them 
for the upcoming analysis. We then make the following ansatz for the compensating $\zeta_{1,2}$-parameters:
\begin{align}
  \label{eq:zetaansatz}
  \zeta_1(\eta) &= a_1 v_\mu(\eta) \gamma^\mu \epsilon_2 + b_1 \gamma^\mu \left(\mathcal{D}(\eta)v(\eta)\right)_\mu \epsilon_2 \,, \nonumber \\
  \zeta_2(\eta) &= a_2 v_\mu(\eta) \gamma^\mu \epsilon_1 + b_2 \gamma^\mu \left(\mathcal{D}(\eta)v(\eta)\right)_\mu \epsilon_1 \,,
\end{align}
where we again kept terms that are zero on-shell, up to the order of derivatives that will be sufficient for the discussion in section \ref{sec:critpoints}. The coefficients $a_1, \cdots, b_2$ depend on $\eta$ and are fixed by requiring that
\begin{align}
 & \gamma^\mu \delta \psi_{\mu}^-(\eta) + \gamma^\mu \left(\bar{D}_\mu \zeta_1(\eta) + \frac{1}{2\ell} \gamma_\mu \zeta_1(\eta)\right) = 0 \,, \nonumber \\
 & \delta\left(\bar{D}^\mu \psi_\mu^-(\eta)\right) + \bar{D}^\mu \left(\bar{D}_\mu \zeta_1(\eta) + \frac{1}{2\ell} \gamma_\mu \zeta_1(\eta)\right) = 0 \,, \nonumber \\
 & \gamma^\mu \delta \psi_{\mu}^+(\eta) + \gamma^\mu \left(\bar{D}_\mu \zeta_2(\eta) - \frac{1}{2\ell} \gamma_\mu \zeta_2(\eta)\right) = 0 \,, \nonumber \\
 & \delta\left(\bar{D}^\mu \psi_\mu^+(\eta)\right) + \bar{D}^\mu \left(\bar{D}_\mu \zeta_2(\eta) - \frac{1}{2\ell} \gamma_\mu \zeta_2(\eta)\right) = 0 \,, 
\end{align}
with $(\mathcal{D}(\eta)^2v)_\mu = 0$. One finds
\begin{align}
  \label{eq:compenszetas}
  \zeta_1(\eta) &= \frac{\ell}{2} \frac{\eta}{(\eta-1)} v_\mu(\eta) \gamma^\mu \epsilon_2 - 
  \frac{\ell^2}{2} \frac{\eta}{(\eta-1)^2} \gamma^\mu \left(\mathcal{D}(\eta)v(\eta)\right)_\mu \epsilon_2 \,, \nonumber \\
\zeta_2(\eta) &= \frac{\ell}{2} \frac{\eta}{(\eta+1)} v_\mu(\eta) \gamma^\mu \epsilon_1 + 
\frac{\ell^2}{2} \frac{\eta}{(\eta+1)^2} \gamma^\mu \left(\mathcal{D}(\eta)v(\eta)\right)_\mu \epsilon_1 \,. 
\end{align}
Finally, we note that the condition $\bar{\nabla}^\mu v_\mu = 0$ is preserved by the gauge fixed supersymmetry transformation rules of (\ref{eq:linsusys}), so no compensating transformation is required for $\delta v_\mu(\eta)$.

\subsubsection{Multiplet structure}

The supersymmetry transformation rules of the modes of the massive multiplets can now be obtained, by adding the compensating diffeomorphisms and fermionic $\zeta$-symmetry of the previous subsection to the 
rules of equations (\ref{eq:linsusys}). One obtains
\begin{align}
  \label{eq:susymassive}
  \delta h_{\mu\nu}(\eta_{1,2}) &= \left(2 + \frac{\eta_{1,2}}{\eta_{1,2}+1} \right) \bar{\epsilon}_1 \gamma_{(\mu} \psi_{\nu)}^-(\eta_{1,2}) + 2 \ell \frac{\eta_{1,2}}{\eta_{1,2}+1} \bar{\epsilon}_1 \bar{\nabla}_{(\mu} \psi_{\nu)}^-(\eta_{1,2}) \nonumber \\ & \qquad + \left(2 + \frac{\eta_{1,2}}{\eta_{1,2}-1} \right) \bar{\epsilon}_2 \gamma_{(\mu} \psi_{\nu)}^+(\eta_{1,2}) - 2 \ell \frac{\eta_{1,2}}{\eta_{1,2}-1} \bar{\epsilon}_2 \bar{\nabla}_{(\mu} \psi_{\nu)}^+(\eta_{1,2}) \,, \nonumber \\
  \delta \psi_{\mu}^-(\eta_{1,2}) &= -\frac14 \gamma^{\rho\sigma} \bar{\nabla}_\rho h_{\mu \sigma}(\eta_{1,2}) \epsilon_1 + 
  \frac{1}{4\ell} h_{\mu\nu}(\eta_{1,2}) \gamma^\nu \epsilon_1 - 
  \frac12 \frac{\eta_{1,2}\ell}{(1-\eta_{1,2})} \left(\bar{D}_\mu v_\nu(\eta_{1,2})\right) \gamma^\nu \epsilon_2 \nonumber \\ 
  & \qquad - \frac12 \frac{\eta_{1,2}}{(1-\eta_{1,2})} v_\mu(\eta_{1,2}) \epsilon_2 + \frac12 v_\nu(\eta_{1,2}) \gamma^\nu \gamma_\mu \epsilon_2 \,, \nonumber \\
  \delta \psi_{\mu}^+(\eta_{1,2}) &= -\frac14 \gamma^{\rho\sigma} \bar{\nabla}_\rho h_{\mu \sigma}(\eta_{1,2}) \epsilon_2 - 
  \frac{1}{4\ell} h_{\mu\nu}(\eta_{1,2}) \gamma^\nu \epsilon_2 + 
  \frac12 \frac{\eta_{1,2}\ell}{(1+\eta_{1,2})} \left(\bar{D}_\mu v_\nu(\eta_{1,2})\right) \gamma^\nu \epsilon_1 \nonumber \\ 
  & \qquad - \frac12 \frac{\eta_{1,2}}{(1+\eta_{1,2})} v_\mu(\eta_{1,2}) \epsilon_1 - \frac12 v_\nu(\eta_{1,2}) \gamma^\nu \gamma_\mu \epsilon_1 \,, \nonumber \\
  \delta v_\mu(\eta_{1,2}) &= -\bar{\epsilon}_1 \slashed{\bar{D}} \psi_{\mu}^+(\eta_{1,2}) + 
  \frac{1}{2\ell} \bar{\epsilon}_1 \psi_{\mu}^+(\eta_{1,2}) + \bar{\epsilon}_2 \slashed{\bar{D}} \psi_{\mu}^-(\eta_{1,2}) + \frac{1}{2\ell} \bar{\epsilon}_2 \psi_{\mu}^-(\eta_{1,2}) \,.
\end{align}
Note that in order to obtain these transformation rules, we have assumed that we are working away from any critical points, i.e. that all modes obey first order 
equations of motion. The above transformation rules should be viewed as {\it solution generating transformations}, in the sense that plugging in solutions of the linearized 
field equations in them, leads to new solutions. One can indeed check that $\delta h_{\mu\nu}(\eta_{1,2})$, $\delta \psi_\mu^\pm(\eta_{1,2})$, $\delta v_\mu(\eta_{1,2})$ obey the correct linearized field equations. 

Since $h_{\mu\nu}(1) = h_{\mu\nu}^L$ and $\psi_\mu^-(1) = \psi_\mu^L$, one can find the supersymmetry transformations of the massless $\{h_{\mu\nu}^L, \psi_\mu^L\}$ multiplet by setting $\eta_{1,2}=1$ in (\ref{eq:susymassive}). Some of the terms in the above transformation rules diverge when setting $\eta_{1,2} = 1$. These terms however involve the modes $v_\mu(1)$ and $\psi_{\mu}^+(1)$ and these can be consistently truncated by virtue of the equations of motion of $h^L_{\mu\nu}$ and $\psi_\mu^L$. One then finds that $h^L_{\mu\nu}$ and $\psi_\mu^L$ only transform into each other via the $\epsilon_1$-supersymmetry according to
\begin{align}
  \label{eq:susymasslessleft}
  \delta h_{\mu\nu}^L &= \frac52 \bar{\epsilon}_1 \gamma_{(\mu} \psi_{\nu)}^L + \ell \bar{\epsilon}_1 \bar{\nabla}_{(\mu} \psi_{\nu)}^L \,, \nonumber \\
  \delta \psi_{\mu}^L &= -\frac14 \gamma^{\rho\sigma} \bar{\nabla}_\rho h^L_{\mu \sigma} \epsilon_1 + \frac{1}{4\ell} h_{\mu\nu}^L \gamma^\nu \epsilon_1 \,.
\end{align}
The transformations of the other massless multiplet $\{h^R_{\mu\nu}, \psi_\mu^R\}$ are then found in a similar way by setting $\eta_{1,2}=-1$ and 
truncating $\psi_\mu^-(-1)$ and $v_\mu(-1)$: 
\begin{align}
  \label{eq:susymasslessright}
  \delta h_{\mu\nu}^R &= \frac52 \bar{\epsilon}_2 \gamma_{(\mu} \psi_{\nu)}^R - \ell \bar{\epsilon}_2 \bar{\nabla}_{(\mu} \psi_{\nu)}^R \,, \nonumber \\
\delta \psi_{\mu}^R &= -\frac14 \gamma^{\rho\sigma} \bar{\nabla}_\rho h^R_{\mu \sigma} \epsilon_2 - \frac{1}{4\ell} h_{\mu\nu}^R \gamma^\nu \epsilon_2 \,.
\end{align}
The modes in these multiplets transform into each other via the two supersymmetries $\epsilon_{1,2}$ in a way that can be summarized in the following diagrams:
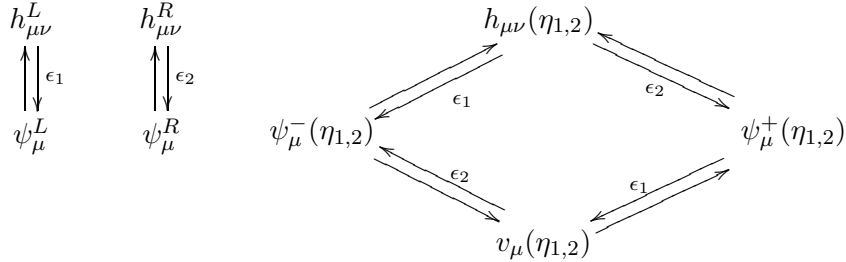
\begin{displaymath}
    \xymatrix{
    h_{\mu\nu}^L\ar@<.5ex>[d]^{\epsilon_1}  \\
      \psi_\mu^L\ar@<.5ex>[u]
        }\qquad
\xymatrix{
    h_{\mu\nu}^R\ar@<.5ex>[d]^{\epsilon_2}  \\
      \psi_\mu^R\ar@<.5ex>[u]
        }\qquad
    \xymatrix{
                                  & {}\quad h_{\mu\nu}(\eta_{1,2})\ \ \ \ \ar@<1ex>[dl]^-{\ \ \epsilon_1}\ar[dr]_-{\ \epsilon_2} &   \\
        \psi_{\mu}^-(\eta_{1,2})\ar[dr]^-{\ \ \epsilon_2}\ar[ur]   &                           &{}\quad \psi_{\mu}^+(\eta_{1,2})\ar[dl]\ar@<-1ex>[ul]\\
                                   & v_\mu(\eta_{1,2})\ar@<-1ex>[ur]^-{\epsilon_1\ \ }\ar@<-1ex>[ul]               &  }
\end{displaymath}

\section{Spectra at critical points} \label{sec:critpoints}

There are 4 separate special cases for which the parameters are such that some of the modes discussed in the previous section coincide. At these critical points in parameter space new logarithmic modes appear. We will now list these 4 cases and discuss the spectrum for each of them in turn.
\begin{itemize}
\item {\underline{Case 1}:} $\eta_1 = \pm 1$, $|\eta_2| \neq 1$ or $\eta_2 = \pm 1$, $|\eta_1| \neq 1$. Let us consider the case $\eta_1 = 1$, $|\eta_2| \neq 1$. This corresponds to a choice of parameters obeying
  \begin{equation}
    \label{eq:case1pars}
    \sigma = \frac{1}{\ell \mu} - \frac{1}{2 \ell^2 m^2} \qquad \mathrm{or} \qquad 2 \mu \sigma \ell^2 m^2 = 2 \ell m^2 - \mu \,.
  \end{equation}
 Here, the gauge fixed equations of motion assume the form
  \begin{align}
    & \left(\mathcal{D}(1)^2\mathcal{D}(-1)\mathcal{D}(\eta_2)h\right)_{\mu\nu}=0\,, \nonumber \\
    & \left( \slashed{\bar{D}} + \frac{1}{2\ell} \right)^2 \left( \slashed{\bar{D}} - \frac{1}{2 \ell} + \frac{1}{\eta_2 \ell} \right) \psi_{\mu\, 1} = 0 \,, \nonumber \\
  & \left( \slashed{\bar{D}} - \frac{1}{2\ell} \right) \left( \slashed{\bar{D}} + \frac{3}{2 \ell}  \right) \left( \slashed{\bar{D}} + \frac{1}{2\ell} + \frac{1}{\eta_2 \ell} \right) \psi_{\mu\, 2} = 0 \,, \nonumber \\
    & \left(\mathcal{D}(1) \mathcal{D}(\eta_2) v \right)_\mu = 0 \,.
  \end{align}
From this one infers that the spectrum consists of one massless multiplet $\{h_{\mu\nu}^R, \psi_\mu^R\}$, one massive 
multiplet $\{h_{\mu\nu}(\eta_2), \psi_{\mu}^\pm(\eta_2), v_\mu(\eta_2)\}$ and one 
`log multiplet' $\{h^{\log L}_{\mu\nu}$, $\psi_\mu^{\log L}$, $\psi_{\mu}^+(1)$, $v_\mu(1)$, $h_{\mu\nu}^L$, $\psi_\mu^L\}$. The modes of the log multiplet obey the 
following equations:
\begin{align}
  \label{eq:logdefeqs}
  & \left(\mathcal{D}(1)^2 h^{\log L}\right)_{\mu\nu} = 0 \quad \mathrm{but} \quad \left(\mathcal{D}(1)h^{\log L}\right)_{\mu\nu} \neq 0  \,, \nonumber \\
  & \left(\slashed{\bar{D}} + \frac{1}{2\ell} \right)^2 \psi_\mu^{\log L} = 0 \quad \mathrm{but} \quad \left(\slashed{\bar{D}} + \frac{1}{2\ell} \right) \psi_\mu^{\log L} \neq 0\,, \nonumber \\
  & \left( \slashed{\bar{D}} + \frac{3}{2 \ell}  \right) \psi_{\mu}^+(1) = 0 \,, \qquad \left(\mathcal{D}(1)v(1)\right)_\mu = 0\,.
\end{align}
The log modes $h^{\log L}_{\mu\nu}$, $\psi_\mu^{\log L}$ defined in this way are only determined up to the addition of massless modes $h_{\mu\nu}^L$, $\psi_\mu^L$. This is why we include them in the 
log multiplet, even though $h_{\mu\nu}^L$ and $\psi_\mu^L$ transform into each other under the $\epsilon_1$-supersymmetry. That these massless modes belong to the `log' multiplet is also evident from the fact that $(h_{\mu\nu}^{\log L}, h_{\mu\nu}^L)$ and $(\psi_\mu^{\log L}, \psi_\mu^L)$ taken together form an indecomposable, non-diagonalizable representation of the Hamiltonian, that takes the form of a rank-2 Jordan cell \cite{Becker:2009mk,Bergshoeff:2011ri,Grumiller:2013at,Grumiller:2008qz,Grumiller:2009mw,Grumiller:2009sn,Grumiller:2010tj}. 

In order to get solution generating supersymmetry transformation rules, we again have to take into account compensating transformations. These can however be 
easily derived from the $\eta \rightarrow 1$ limits of the
compensating diffeomorphism and $\zeta$-transformation parameters given in equations (\ref{eq:compdiffeoresult}), (\ref{eq:compenszetas}). Adding for instance a 
compensating diffeomorphism with parameter $\xi_\mu(\eta)$ given in (\ref{eq:compdiffeoresult}), we get the following rule
\begin{align}
  \label{eq:deltahthreeders}
  \delta h_{\mu\nu}(\eta) &= \left(2 + \frac{f(\eta)}{\ell}\right) \bar{\epsilon}_1 \gamma_{(\mu} \psi_{\nu)}^-(\eta) + \frac{f(\eta)^2}{\ell} \bar{\epsilon}_1 \gamma_{(\mu} \left(\slashed{\bar{D}} - \frac{1}{2\ell} + \frac{1}{\eta \ell}\right) \psi_{\nu)}^-(\eta) \nonumber \\ & \quad  + 2 f(\eta) \bar{\epsilon}_1 \bar{\nabla}_{(\mu} \psi_{\nu)}^- (\eta) + 2 f(\eta)^2 \bar{\epsilon}_1 \bar{\nabla}_{(\mu} \left(\slashed{\bar{D}} - \frac{1}{2\ell} + \frac{1}{\eta \ell}\right) \psi_{\nu)}^-(\eta) + \cdots \,,
\end{align}
where the $\cdots$ contain terms that involve $\left(\slashed{\bar{D}} - \frac{1}{2\ell} + \frac{1}{\eta \ell}\right)^2 \psi_{\mu}^-(\eta)$ and an $\epsilon_2$ 
transformation $2 \bar{\epsilon}_2 \gamma_{(\mu} \psi_{\mu)}^+(\eta)$ together with compensator terms for this $\epsilon_2$ transformation that involve $g(\eta)$ defined
in (\ref{fg}). We can 
then consider this rule in the $\eta \rightarrow 1$ limit and see what happens when one plugs in $\psi_\nu^L$ or $\psi_\nu^{\mathrm{log}\, L}$ for the gravitino 
mode $\psi_{\nu}^-(\eta)$. Since $g(\eta)$ diverges in the $\eta \rightarrow 1$ limit, one finds that there exists no compensator for the $\epsilon_2$ supersymmetry. The 
latter can thus not be used as a solution generating symmetry. In the $\eta \rightarrow 1$ limit, the terms involving 
$\left(\slashed{\bar{D}} - \frac{1}{2\ell} + \frac{1}{\eta \ell}\right)^2 \psi_{\mu}^-(\eta)$ always vanish, regardless of whether $\psi_\mu^-(\eta)$ corresponds to 
$\psi_\mu^L$ or $\psi_\mu^{\mathrm{log}\, L}$. 
Equation (\ref{eq:deltahthreeders}) then reduces to
\begin{align}
  \label{eq:deltahthreedersfinal}
  \delta \left(\lim_{\eta \rightarrow 1} h_{\mu\nu}(\eta)\right) &= \frac52 \bar{\epsilon}_1 \gamma_{(\mu} \left( \lim_{\eta \rightarrow 1} \psi_{\nu)}^-(\eta)\right) + \ell \bar{\epsilon}_1 \bar{\nabla}_{(\mu} \left( \lim_{\eta \rightarrow 1} \psi_{\nu)}^- (\eta)\right) \nonumber \\ & \quad  + \frac{\ell}{4} \bar{\epsilon}_1 \gamma_{(\mu} \left(\slashed{\bar{D}} + \frac{1}{2\ell} \right) \left( \lim_{\eta \rightarrow 1} \psi_{\nu)}^-(\eta)\right) \nonumber \\ & \quad  + \frac{\ell^2}{2} \bar{\epsilon}_1 \bar{\nabla}_{(\mu} \left(\slashed{\bar{D}} + \frac{1}{2\ell} \right) \left( \lim_{\eta \rightarrow 1} \psi_{\nu)}^-(\eta)\right)  \,.
\end{align}
This transformation rule obeys the gauge fixing conditions for the metric perturbation and thus indeed contains the correct compensating transformations. Replacing $\lim_{\eta \rightarrow 1} \psi_{\mu}^-(\eta)$ by $\psi_\mu^L$, we find that the last two terms of (\ref{eq:deltahthreedersfinal}) vanish, while the first two are annihilated by $\mathcal{D}(1)$. We can thus conclude that an $\epsilon_1$ transformation can be used to generate a massless graviton solution from a massless gravitino one:
\begin{align}
  \delta h_{\mu\nu}^L &= \frac52 \bar{\epsilon}_1 \gamma_{(\mu} \psi_{\nu)}^L + \ell \bar{\epsilon}_1 \bar{\nabla}_{(\mu} \psi_{\nu)}^L \,.
\end{align}
If one replaces $\lim_{\eta \rightarrow 1} \psi_{\mu}^-(\eta)$ in (\ref{eq:deltahthreedersfinal}) by $\psi_\mu^{\mathrm{log}\, L}$, one finds that the 
right hand side is annihilated by $\mathcal{D}(1)^2$. In this way, an $\epsilon_1$ transformation generates a graviton log mode from a gravitino one:
\begin{align}
  \delta h_{\mu\nu}^{\mathrm{log}\, L} &= \frac52 \bar{\epsilon}_1 \gamma_{(\mu} \psi_{\nu)}^{\mathrm{log}\, L} + 
  \ell \bar{\epsilon}_1 \bar{\nabla}_{(\mu} \psi_{\nu)}^{\mathrm{log}\, L} + \frac{\ell}{4} \bar{\epsilon}_1 \gamma_{(\mu} \left(\slashed{\bar{D}} + 
  \frac{1}{2\ell} \right) \psi_{\nu)}^{\mathrm{log}\, L} \nonumber \\ & \quad   + 
  \frac{\ell^2}{2} \bar{\epsilon}_1 \bar{\nabla}_{(\mu} \left(\slashed{\bar{D}} + \frac{1}{2\ell} \right) \psi_{\nu)}^{\mathrm{log}\, L}  \,.
\end{align}

A similar discussion holds for $\delta \psi_\mu^-(\eta)$. In this case, we found in subsection \ref{sssec:compensatingtrafos} that only the $\epsilon_2$ transformation needed a $\zeta_1$-compensator. The compensator derived in (\ref{eq:compenszetas}) however diverges in the $\eta\rightarrow 1$ limit, implying that the $\epsilon_2$-supersymmetry cannot be used as a solution generating symmetry. In the limit $\eta \rightarrow 1$, $\delta \psi_\mu^-(\eta)$ thus only contains the $\epsilon_1$-supersymmetry 
\begin{equation} \label{eq:deltapsilogold}
  \delta \left( \lim_{\eta \rightarrow 1} \psi_{\mu}^-(\eta)\right) = -\frac14 \gamma^{\rho\sigma} \bar{\nabla}_\rho \left( \lim_{\eta \rightarrow 1} h_{\mu\sigma}(\eta)\right) \epsilon_1 + \frac{1}{4\ell} \left( \lim_{\eta \rightarrow 1} h_{\mu\nu}(\eta)\right) \gamma^\nu \epsilon_1 \,,
\end{equation}
Replacing $\lim_{\eta \rightarrow 1} h_{\mu\nu}(\eta)$ by $h_{\mu\nu}^L$, one finds that the right-hand-side is annihilated by $(\slashed{\bar{D}} + \frac{1}{2\ell})$ and hence that a massless gravitino mode is generated from a massless graviton one
\begin{equation}
  \delta \psi_{\mu}^L = -\frac14 \gamma^{\rho\sigma} \bar{\nabla}_\rho h_{\mu\sigma}^L \epsilon_1 + \frac{1}{4\ell} h_{\mu\nu}^L \gamma^\nu \epsilon_1 \,.
\end{equation}
Replacing $\lim_{\eta \rightarrow 1} h_{\mu\nu}(\eta)$ by $h_{\mu\nu}^{\mathrm{log}\, L}$, one finds that the right-hand-side of (\ref{eq:deltapsilogold}) is annihilated by $(\slashed{\bar{D}} + \frac{1}{2\ell})^2$ and hence that a logarithmic gravitino mode is generated from a logarithmic graviton one
\begin{equation} \label{eq:deltapsilog}
  \delta_{\epsilon_1} \psi_{\mu}^{\log L} = -\frac14 \gamma^{\rho\sigma} \bar{\nabla}_\rho h^{\log L}_{\mu\sigma} \epsilon_1 + \frac{1}{4\ell} h_{\mu\nu}^{\log L} \gamma^\nu \epsilon_1 \,.
\end{equation}
Finally, for the massive modes $\psi^{+}_{\mu}(1)$ and $v_{\mu}(1)$ one similarly reasons to obtain the following solution generating supersymmetry transformation rules\
\begin{align}
  \delta \psi_{\mu}^{+}(1) &= -\frac14 \gamma^{\rho\sigma} \bar{\nabla}_\rho h_{\mu \sigma}^{\mathrm{log \, } L} \epsilon_2 - \frac{1}{4\ell} h_{\mu\nu}^{\mathrm{log \,} L} \gamma^\nu \epsilon_2 + \frac{\ell}{4} \left(\bar{D}_\mu v_\nu(1)\right) \gamma^\nu \epsilon_1 - \frac14  v_\mu(1) \epsilon_1 \nonumber \\ & \qquad - \frac12 v_\nu(1) \gamma^\nu \gamma_\mu \epsilon_1 \,, \nonumber \\
  \delta v_\mu(1) &= -\bar{\epsilon}_1 \bar{\slashed{D}} \psi_{\mu}^+(1) + \frac{1}{2\ell} \bar{\epsilon}_1 \psi_{\mu}^+(1) + \bar{\epsilon}_2 \left(\bar{\slashed{D}} +\frac{1}{2\ell}\right)\psi_{\mu}^{\mathrm{log \,} L}\,.  
\end{align}

The transformation properties of the various modes in the log multiplet can be summarized in the following diagram:
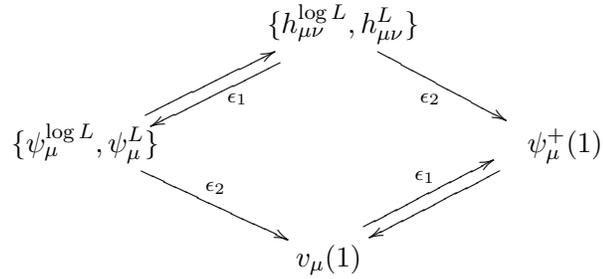
\begin{displaymath}
    \xymatrix{
                                  & {}\quad \{h_{\mu\nu}^{\mathrm{log \,} L},h_{\mu\nu}^L\} \ar@<1ex>[dl]\ar[dr]_{\epsilon_2} &   \\
        \{\psi_\mu^{\mathrm{log \,} L},\psi_\mu^L\}\ar[dr]^{\epsilon_2}\ar[ur]_{\ \ \ \epsilon_1}   &                          &{}\quad \psi_{\mu}^{+}(1)\ar[dl] \\
                                   & v_\mu(1)\ar@<1ex>[ur]^{\epsilon_1}              &  }
\end{displaymath}

\item {\underline{Case 2:}} $\eta_1 = \pm 1$, $\eta_2 = \mp 1$. Let us consider the case $\eta_1 = 1$, $\eta_2 =- 1$ without 
loss of generality. This corresponds to a choice of parameters for which
  \begin{equation}
    \label{eq:case2pars}
    \mu^{-1} = 0 \,.
  \end{equation} 
Here, the gravitational Chern-Simons term and its supersymmetrization are thus absent and one recovers the 
$\mathcal{N}=(1,1)$ supersymmetrization of NMG \cite{Bergshoeff:2009hq}. The four-dimensional counterpart of this case was studied in \cite{Lu:2011mw}.
Here, the equations of motion take the form:
  \begin{align}
    & \left(\mathcal{D}(1)^2\mathcal{D}(-1)^2h\right)_{\mu\nu}=0\,, \nonumber \\
    & \left( \slashed{\bar{D}} + \frac{1}{2\ell} \right)^2 \left( \slashed{\bar{D}} - \frac{3}{2 \ell} \right) \psi_{\mu\, 1} = 0 \,, \nonumber \\
  & \left( \slashed{\bar{D}} - \frac{1}{2\ell} \right)^2 \left( \slashed{\bar{D}} + \frac{3}{2 \ell}  \right)  \psi_{\mu\, 2} = 0 \,, \nonumber \\
    & \left(\mathcal{D}(1) \mathcal{D}(-1) v \right)_\mu = 0 \,.
  \end{align}
In this case, there are two log multiplets of the type we encountered in the previous case: $\{h^{\log L}_{\mu\nu}, \psi_\mu^{\log L}, \psi_{\mu}^{+}(1), v_\mu(1),h_{\mu\nu}^L, \psi_\mu^L\}$ and $\{h^{\log R}_{\mu\nu}, \psi_\mu^{\log R}, \psi_{\mu}^{-}(-1), v_\mu(-1),h_{\mu\nu}^R, \psi_\mu^R\}$. Their supersymmetry transformation rules can be determined as in the previous case. They can be summarized in the following diagrams:
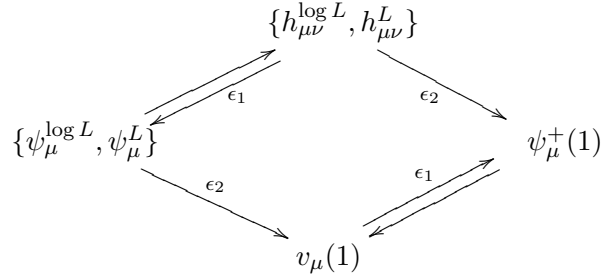
\begin{displaymath}
    \xymatrix{
                                  & {}\quad \{h_{\mu\nu}^{\mathrm{log \, } L},h_{\mu\nu}^L\} \ar@<1ex>[dl]\ar[dr]_{\epsilon_2} &   \\
        \{\psi_\mu^{\mathrm{log \,} L},\psi_\mu^L\}\ar[dr]^{\epsilon_2}\ar[ur]_{\ \ \ \epsilon_1}   &                          &{}\quad \psi_{\mu}^{+}(1)\ar[dl] \\
                                   & v_\mu(1)\ar@<1ex>[ur]^{\epsilon_1}              &  }
\end{displaymath}
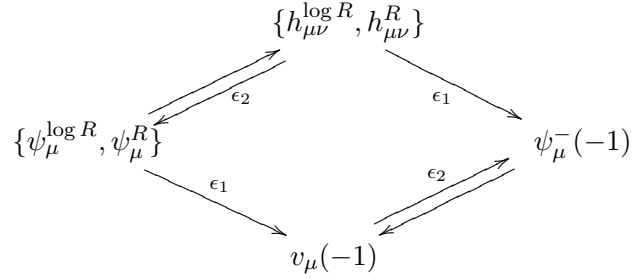
\begin{displaymath}
    \xymatrix{
                                  & {}\quad \{h_{\mu\nu}^{\mathrm{log \,} R},h_{\mu\nu}^R\} \ar@<1ex>[dl]\ar[dr]_{\epsilon_1} &   \\
        \{\psi_\mu^{\mathrm{log \,}R},\psi_\mu^R\}\ar[dr]^{\epsilon_1}\ar[ur]_{\ \ \ \epsilon_2}   &                          &{}\quad \psi_{\mu}^{-}(-1)\ar[dl] \\
                                   & v_\mu(-1)\ar@<1ex>[ur]^{\epsilon_2}              &  }
\end{displaymath}

\item {\underline{Case 3}:} $\eta_1 = \pm 1$, $\eta_2 = \pm 1$. Let us choose $\eta_1 = \eta_2 = 1$ without loss of generality. In this case, the parameters have to obey
  \begin{equation}
    \label{eq:case3pars}
    \sigma \ell^2 m^2 = \frac32 \qquad \mathrm{and} \qquad m^2 \ell = 2 \mu \,.
  \end{equation}
Here, the equations of motion become:
  \begin{align}
    & \left(\mathcal{D}(1)^3\mathcal{D}(-1)h\right)_{\mu\nu}=0\,, \nonumber \\
    & \left( \slashed{\bar{D}} + \frac{1}{2\ell} \right)^3 \psi_{\mu\, 1} = 0 \,, \nonumber \\
  & \left( \slashed{\bar{D}} - \frac{1}{2\ell} \right) \left( \slashed{\bar{D}} + \frac{3}{2 \ell}  \right)^2  \psi_{\mu\, 2} = 0 \,, \nonumber \\
    & \left(\mathcal{D}(1)^2  v \right)_\mu = 0 \,.
  \end{align}
The spectrum thus consists of one massless multiplet $\{h_{\mu\nu}^R, \psi_\mu^R\}$ and one `log$^2$ 
multiplet' $\{h^{\log^2 L}_{\mu\nu}, h^{\log L}_{\mu\nu},\psi^{\log^2 L}_\mu, \psi_\mu^{\log L}, \psi_{\mu}^{\log\, +}(1), \psi_{\mu}^{+}(1), v_\mu^{\log}(1), v_\mu(1),h_{\mu\nu}^L, \psi_\mu^L\}$. This multiplet thus consists of log modes that are defined as previously, as well as log$^2$ modes that obey
\begin{align}
  \label{eq:logdefeqs2}
  & \left(\mathcal{D}(1)^3 h^{\log^2 L}\right)_{\mu\nu} = 0 \quad \mathrm{but} \quad \left(\mathcal{D}(1)^2h^{\log^2 L}\right)_{\mu\nu} \neq 0  \,, \nonumber \\
  & \left(\slashed{\bar{D}} + \frac{1}{2\ell} \right)^3 \psi_\mu^{\log^2 L} = 0 \quad \mathrm{but} 
  \quad \left(\slashed{\bar{D}} + \frac{1}{2\ell} \right)^2 \psi_\mu^{\log^2 L} \neq 0\,, \nonumber \\
  & \left( \slashed{\bar{D}} + \frac{3}{2 \ell}  \right)^2 \psi_{\mu}^{\log\, +}(1) = 0 \quad \mathrm{but} \quad \left( \slashed{\bar{D}} + \frac{3}{2 \ell}  \right) \psi_{\mu}^{\log\, +}(1) \neq 0  \,, \nonumber \\ & \left(\mathcal{D}(1)^2v^{\log}(1)\right)_\mu = 0 \quad \mathrm{but} \quad \left(\mathcal{D}(1)v^{\log}(1)\right)_\mu \neq 0 \,.
\end{align}
The log$^2$ modes are only defined up to the addition of massless modes $h_{\mu\nu}^L$, $\psi_\mu^L$ and log 
modes $h_{\mu\nu}^{\log L}$, $\psi_\mu^{\log L}$. In this case, 
$(h_{\mu\nu}^{\log^2 L},h_{\mu\nu}^{\log L}, h_{\mu\nu}^L)$ and $(\psi_\mu^{\log^2 L},\psi_\mu^{\log L}, \psi_\mu^L)$ taken together 
form an indecomposable, non-diagonalizable representation of the Hamiltonian, that takes the form of a rank-3 Jordan cell \cite{Grumiller:2010tj}. Solution 
generating supersymmetry transformation rules can be obtained as before. As in the previous two cases one finds that some of the supersymmetry transformations 
are not invertible, due to the fact that the necessary compensating transformations diverge for $\eta=1$. 

Solution generating supersymmetry transformations can be obtained using the compensating diffeomorphisms (\ref{eq:compdiffeoresult}) and fermionic symmetry (\ref{eq:compenszetas}), following the reasoning outlined in case 1. Here, all higher derivative terms given in (\ref{eq:compdiffeoresult}) and (\ref{eq:compenszetas}) contribute to the compensating transformations. The transformation properties of the various modes in the log multiplet can be summarized in the following diagram:
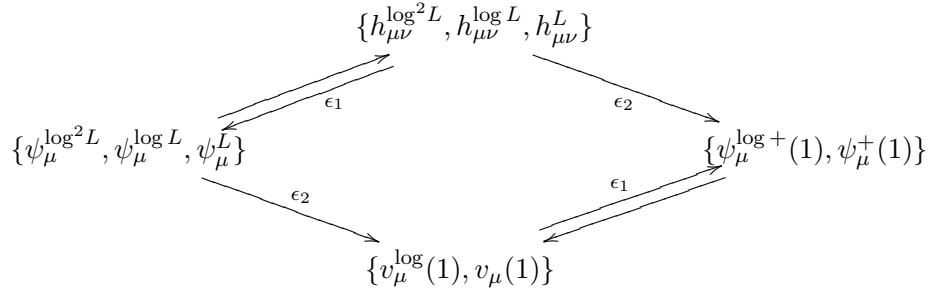
\begin{displaymath}
    \xymatrix{
                                  & {}\quad \{h_{\mu\nu}^{\mathrm{log}^2 L}, h_{\mu\nu}^{\mathrm{log \,}L},h_{\mu\nu}^L\} \ar@<1ex>[dl]\ar[dr]_{\epsilon_2} &   \\
        \{\psi_\mu^{\mathrm{log}^2 L}, \psi_\mu^{\mathrm{log \,} L},\psi_\mu^L\}\ar[dr]^{\epsilon_2}\ar[ur]_{\ \ \ \ \ \epsilon_1}   &                          &{}\quad \{\psi_\mu^{\mathrm{log}\, +}(1), \psi_{\mu}^{+}(1)\}\ar[dl] \\
                                   & \{v_\mu^{\mathrm{log}}(1), v_\mu(1)\}\ar@<1ex>[ur]^{\epsilon_1}              &  }
\end{displaymath}
\item {\underline{Case 4:}} $\eta_1 = \eta_2 = \eta$, $|\eta| \neq 1$. This case is obtained when the parameters obey the following constraint
  \begin{equation}
    \label{eq:case4pars}
    \mu^2 = \frac{m^4 \ell^2}{4 \sigma \ell^2 m^2 - 2} \,.
  \end{equation}
In this case, the equations of motion are:
  \begin{align}
    & \left(\mathcal{D}(1)\mathcal{D}(-1) \mathcal{D}(\eta)^2 h\right)_{\mu\nu}=0\,, \nonumber \\
    & \left( \slashed{\bar{D}} + \frac{1}{2\ell} \right) \left( \slashed{\bar{D}} - \frac{1}{2 \ell} + \frac{1}{\eta \ell} \right)^2 \psi_{\mu\, 1} = 0 \,, \nonumber \\
  & \left( \slashed{\bar{D}} - \frac{1}{2\ell} \right) \left( \slashed{\bar{D}} + \frac{1}{2 \ell} + \frac{1}{\eta \ell}  \right)^2  \psi_{\mu\, 2} = 0 \,, \nonumber \\
    & \left(\mathcal{D}(\eta)^2 v \right)_\mu = 0 \,.
  \end{align}
The spectrum consists of 2 massless multiplets $\{h_{\mu\nu}^{L/R}, \psi_\mu^{L/R}\}$ and one `massive log' multiplet $\{h_{\mu\nu}^{\log}(\eta), h_{\mu\nu}(\eta), \psi_{\mu}^{\log\, +}(\eta), \psi_{\mu}^{\log\, -}(\eta), \psi_{\mu}^+(\eta), \psi_{\mu}^-(\eta), v_\mu^{\log}(\eta), v_\mu(\eta)\}$. The log modes obey:
\begin{align}
  \label{eq:massivelogdefeqs}
  & \left(\mathcal{D}(\eta)^2 h^{\log}(\eta)\right)_{\mu\nu} = 0 \quad \mathrm{but} \quad \left(\mathcal{D}(\eta)h^{\log}(\eta)\right)_{\mu\nu} \neq 0  \,, \nonumber \\
  & \left(\slashed{\bar{D}} \pm \frac{1}{2\ell} + \frac{1}{\eta \ell} \right)^2 \psi_{\mu}^{\log\, \pm}(\eta) = 0 \quad \mathrm{but} \quad \left(\slashed{\bar{D}} \pm \frac{1}{2\ell} + \frac{1}{\eta \ell} \right) \psi_{\mu}^{\log\, \pm}(\eta) \neq 0\,, \nonumber \\
  & \left(\mathcal{D}(\eta)^2 v^{\log}(\eta)\right)_{\mu} = 0 \quad \mathrm{but} \quad \left(\mathcal{D}(\eta)v^{\log}(\eta)\right)_{\mu} \neq 0 \,.
\end{align}
Since none of the compensating transformations diverge for these critical points, one finds that all supersymmetry transformations are invertible and the transformation properties can be summarized in the following diagram
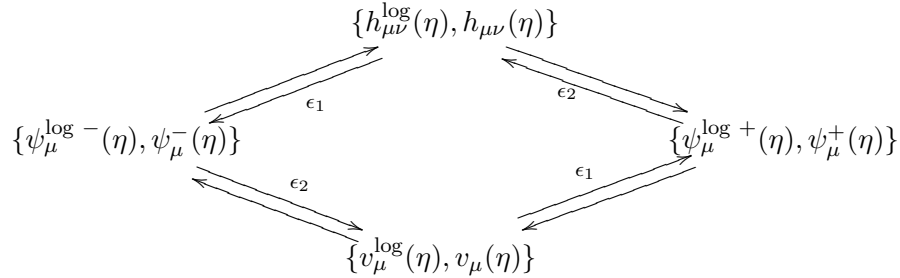
\begin{displaymath}
    \xymatrix{
                                  & {}\quad \{h_{\mu\nu}^{\log}(\eta), h_{\mu\nu}(\eta)\} \ar@<1ex>[dl]^-{\ \ \epsilon_1}\ar[dr]_-{\ \epsilon_2\ \ \ \ } &   \\
        \{\psi_{\mu}^{\log\, -}(\eta), \psi_{\mu}^{-}(\eta)\}\ar[dr]^-{\ \ \epsilon_2}\ar[ur]   &                           &{}\quad \{\psi_{\mu}^{\log\, +}(\eta), \psi_{\mu}^{+}(\eta)\}\ar[dl]\ar@<1ex>[ul]\\
                                   & \{v_\mu^{\log}(\eta), v_\mu(\eta)\}\ar@<1ex>[ur]^-{\epsilon_1\ \ }\ar@<1ex>[ul]               &  }
\end{displaymath}

\end{itemize}

\section{Conclusions and Outlook} \label{sec:concl}

Motivated by obtaining supersymmetric generalizations of the AdS/log CFT correspondence, we have considered three-dimensional $\mathcal{N}=(1,1)$ supersymmetric 
GMG models. First, we have obtained the linearized supersymmetry transformation rules and linearized action, including fermionic terms of $\mathcal{N}=(1,1)$ GMG and 
studied the spectrum of linearized fluctuations and how they assemble themselves in massless and massive spin-2 supermultiplets for generic points in parameter space. 
We have then looked at critical points in parameter space, where some of the massive modes become degenerate with massless or other massive modes and where so-called 
logarithmic modes appear. In particular, we have argued that there are four classes of critical points: one class where there is one supermultiplet containing logarithmic 
modes, along with massless and massive modes, a second class with two such supermultiplets, a third class that contains a supermultiplet with logarithmic and doubly 
logarithmic modes, along with massless and massive modes and finally, a fourth class with a supermultiplet of massive and logarithmic massive modes. For each of these 
classes, we have described the multiplet structure and given the supersymmetry transformation rules that connect the various modes.

The results of this paper can be used as a starting point for several interesting research directions and extensions. For applications to the AdS/log CFT 
correspondence, it would be desirable to have a better understanding of representations of AdS superalgebras, where the AdS Hamiltonian acts in an indecomposable, 
non-diagonalizable fashion on the states. Since such representations are non-unitary, standard classification theorems for unitary representations no longer apply. In this paper, we have encountered several examples, that could be used to better develop the theory of 
such indecomposable representations of AdS superalgebras. As we have seen in this paper and as was observed in \cite{Lu:2011mw}, some of the supersymmetry transformations can be realized in a non-invertible manner in such multiplets. Another interesting direction concerns the application of the holographic dictionary to calculate two- and 
three-point correlators of the boundary stress-energy tensor in critical $\mathcal{N}=(1,1)$ GMG. The results obtained in this way could then be compared with results 
for stress-energy correlators in supersymmetric log CFTs \cite{Khorrami:1998kw,Drichel:2010eb,Pearce:2013bea}. In order to calculate these correlators on the gravitational side, the strategy of \cite{Grumiller:2009mw} could be followed. One then needs explicit solutions for the 
non-normalizable modes, that arise as solutions of the linearized field equations. For the graviton modes, such solutions have been considered in the context of critical 
TMG in \cite{Grumiller:2009mw}. It would therefore be interesting to extend this analysis to the gravitino sector as well. Note that to be able to properly apply the 
holographic dictionary, the action typically needs to be supplemented with boundary terms in order to render the variational principle well-defined. Such boundary 
terms have typically been ignored in the construction of supersymmetric GMG models and their inclusion therefore gives a novel direction for further research. 
Let us note that in the case of three-dimensional Einstein gravity the boundary term that is required for a well-defined variational principle 
coincides with the boundary term obtained by requiring supersymmetry in the presence of a boundary, without imposing boundary conditions on the fields 
\cite{Grumiller:2009dx}. It would be interesting to see whether the reasoning of \cite{Grumiller:2009dx} can be applied to (supersymmetric) GMG as well.

\section{Acknowledgments}

JR acknowledges funding from the Short Term Scientific Mission COST-STSM-MP1210-25197 of the COST MP1210 `The String Theory Universe' network and financial support by the START 
project Y 435-N16 of the Austrian Science Fund (FWF) and by the NCCR SwissMAP, funded by the Swiss National Science Foundation. NSD is partially supported by the Scientific and Technological Research Council of Turkey (T\"ubitak) project 116F137.  
We also thank the Galileo Galilei Institute for Theoretical Physics and the Albert Einstein Institute, Potsdam for hospitality during part of this work. JR thanks Bo\u{g}azi\c{c}i University for hospitality during part of this work.

\appendix

\section{Notation and conventions} \label{app:conv}

When dealing with complex Dirac spinors, we have adopted the conventions of \cite{Alkac:2014hwa}. The Minkowski metric is taken to have mostly plus 
signature: $\eta_{ab} = \mathrm{diag}(-1,1,1)$ and the gamma matrices are taken such that they satisfy the Clifford algebra relation 
$\{\gamma_a, \gamma_b\} = 2 \eta_{ab}\, \mathbb{1}$. Given an irreducible $(2\times 2)$-dimensional Clifford algebra representation, the 
matrices $(\gamma_a)^\dag$, $-(\gamma_a)^T$ and $(\gamma_a)^*$ are similar to the gamma matrices $\gamma_a$. The similarity matrices are $\gamma^0$, the charge 
conjugation matrix $C$ and a matrix $B$ respectively:
\begin{equation}
  \label{eq:defCB}
  (\gamma_a)^\dag = \gamma^0 \gamma_a \gamma^0 \,, \qquad (\gamma_a)^T = - C \gamma_a C^{-1} \,, \qquad (\gamma_a)^* = B \gamma_a B^{-1} \,.
\end{equation}
The charge conjugation matrix $C$ and the matrix $B$ satisfy the following properties:
\begin{equation}
  C C^\dag = \mathbb{1} \,, \qquad C C^* = - \mathbb{1} \,, \qquad C^T = - C \,,
\end{equation}
and
\begin{equation}
  C = \rmi B \gamma^0 \,, \qquad B B^\dag = \mathbb{1} \,, \qquad B B^* = \mathbb{1} \,, \qquad B^T = B \,.
\end{equation}
For Dirac spinors, we use two different definitions of the conjugate spinors:
\begin{equation}
  \label{eq:defconjugates}
  \bar{\epsilon} = \rmi \epsilon^\dag \gamma^0 \,, \qquad \tilde{\epsilon} = \overline{(B \epsilon)^*} \,.
\end{equation}
Majorana spinors are defined as spinors that satisfy the Majorana condition $\epsilon^* = B \epsilon$. For Majorana spinors, one then has that $\bar{\epsilon} = \tilde{\epsilon}$. A Dirac spinor $\epsilon$ can then be split up in real and imaginary parts:
\begin{equation}
  \label{eq:diracsplit}
  \epsilon = \epsilon_1 + \rmi \epsilon_2 \,, \qquad \mathrm{with} \quad \epsilon_1 = \frac12 \left(\epsilon + (B\epsilon)^* \right) \,,\ \  \epsilon_2 = -\frac{\rmi}{2} \left(\epsilon - (B \epsilon)^* \right)\,.
\end{equation}
Here, the real and imaginary parts $\epsilon_1$ and $\epsilon_2$ are Majorana spinors. In this paper, we will denote the real and imaginary parts of a Dirac spinor with a subindex 1 and 2 respectively. We refer to \cite{Alkac:2014hwa} for further properties and useful calculational tips.

%\bibliography{susyAdSLCFT}

\begin{thebibliography}{10}

\bibitem{Gurarie:1993xq}
V.~Gurarie, ``{Logarithmic operators in conformal field theory},''
  \href{http://dx.doi.org/10.1016/0550-3213(93)90528-W}{{\em Nucl. Phys.} {\bf
  B410} (1993)  535--549},
\href{http://arxiv.org/abs/hep-th/9303160}{{\tt arXiv:hep-th/9303160
  [hep-th]}}.
%%CITATION = HEP-TH/9303160;%%.

\bibitem{Flohr:1996ik}
M.~A.~I. Flohr, ``{Two-dimensional turbulence: Yet another conformal field
  theory solution},''
  \href{http://dx.doi.org/10.1016/S0550-3213(96)00563-9}{{\em Nucl. Phys.} {\bf
  B482} (1996)  567--578},
\href{http://arxiv.org/abs/hep-th/9606130}{{\tt arXiv:hep-th/9606130
  [hep-th]}}.
%%CITATION = HEP-TH/9606130;%%.

\bibitem{Grumiller:2013at}
D.~Grumiller, W.~Riedler, J.~Rosseel, and T.~Zojer, ``{Holographic applications
  of logarithmic conformal field theories},''
  \href{http://dx.doi.org/10.1088/1751-8113/46/49/494002}{{\em J. Phys.} {\bf
  A46} (2013)  494002},
\href{http://arxiv.org/abs/1302.0280}{{\tt arXiv:1302.0280 [hep-th]}}.
%%CITATION = ARXIV:1302.0280;%%.

\bibitem{Grumiller:2008qz}
D.~Grumiller and N.~Johansson, ``{Instability in cosmological topologically
  massive gravity at the chiral point},''
  \href{http://dx.doi.org/10.1088/1126-6708/2008/07/134}{{\em JHEP} {\bf 07}
  (2008)  134},
\href{http://arxiv.org/abs/0805.2610}{{\tt arXiv:0805.2610 [hep-th]}}.
%%CITATION = ARXIV:0805.2610;%%.

\bibitem{Bergshoeff:2009hq}
E.~A. Bergshoeff, O.~Hohm, and P.~K. Townsend, ``{Massive Gravity in Three
  Dimensions},'' \href{http://dx.doi.org/10.1103/PhysRevLett.102.201301}{{\em
  Phys. Rev. Lett.} {\bf 102} (2009)  201301},
\href{http://arxiv.org/abs/0901.1766}{{\tt arXiv:0901.1766 [hep-th]}}.
%%CITATION = ARXIV:0901.1766;%%.

\bibitem{Bergshoeff:2009aq}
E.~A. Bergshoeff, O.~Hohm, and P.~K. Townsend, ``{More on Massive 3D
  Gravity},'' \href{http://dx.doi.org/10.1103/PhysRevD.79.124042}{{\em Phys.
  Rev.} {\bf D79} (2009)  124042},
\href{http://arxiv.org/abs/0905.1259}{{\tt arXiv:0905.1259 [hep-th]}}.
%%CITATION = ARXIV:0905.1259;%%.

\bibitem{Lu:2011zk}
H.~Lu and C.~N. Pope, ``{Critical Gravity in Four Dimensions},''
  \href{http://dx.doi.org/10.1103/PhysRevLett.106.181302}{{\em Phys. Rev.
  Lett.} {\bf 106} (2011)  181302},
\href{http://arxiv.org/abs/1101.1971}{{\tt arXiv:1101.1971 [hep-th]}}.
%%CITATION = ARXIV:1101.1971;%%.

\bibitem{Deser:2011xc}
S.~Deser, H.~Liu, H.~Lu, C.~N. Pope, T.~C. Sisman, and B.~Tekin, ``{Critical
  Points of D-Dimensional Extended Gravities},''
  \href{http://dx.doi.org/10.1103/PhysRevD.83.061502}{{\em Phys. Rev.} {\bf
  D83} (2011)  061502},
\href{http://arxiv.org/abs/1101.4009}{{\tt arXiv:1101.4009 [hep-th]}}.
%%CITATION = ARXIV:1101.4009;%%.

\bibitem{Bergshoeff:2011ri}
E.~A. Bergshoeff, O.~Hohm, J.~Rosseel, and P.~K. Townsend, ``{Modes of Log
  Gravity},'' \href{http://dx.doi.org/10.1103/PhysRevD.83.104038}{{\em Phys.
  Rev.} {\bf D83} (2011)  104038},
\href{http://arxiv.org/abs/1102.4091}{{\tt arXiv:1102.4091 [hep-th]}}.
%%CITATION = ARXIV:1102.4091;%%.

\bibitem{Gates:2013tka}
S.~J. Gates, Jr. and K.~Koutrolikos, ``{A dynamical theory for linearized
  massive superspin 3/2},''
  \href{http://dx.doi.org/10.1007/JHEP03(2014)030}{{\em JHEP} {\bf 03} (2014)
  030},
\href{http://arxiv.org/abs/1310.7387}{{\tt arXiv:1310.7387 [hep-th]}}.
%%CITATION = ARXIV:1310.7387;%%.

\bibitem{Lu:2011mw}
H.~Lu, C.~N. Pope, E.~Sezgin, and L.~Wulff, ``{Critical and Non-Critical
  Einstein-Weyl Supergravity},''
  \href{http://dx.doi.org/10.1007/JHEP10(2011)131}{{\em JHEP} {\bf 10} (2011)
  131},
\href{http://arxiv.org/abs/1107.2480}{{\tt arXiv:1107.2480 [hep-th]}}.
%%CITATION = ARXIV:1107.2480;%%.

\bibitem{Deser:1982sw}
S.~Deser and J.~H. Kay, ``{Topologically Massive Supergravity},''
\href{http://dx.doi.org/10.1016/0370-2693(83)90631-7}{{\em Phys. Lett.} {\bf
  B120} (1983)  97--100}.
%%CITATION = PHLTA,B120,97;%%.

\bibitem{Andringa:2009yc}
R.~Andringa, E.~A. Bergshoeff, M.~de~Roo, O.~Hohm, E.~Sezgin, and P.~K.
  Townsend, ``{Massive 3D Supergravity},''
  \href{http://dx.doi.org/10.1088/0264-9381/27/2/025010}{{\em Class. Quant.
  Grav.} {\bf 27} (2010)  025010},
\href{http://arxiv.org/abs/0907.4658}{{\tt arXiv:0907.4658 [hep-th]}}.
%%CITATION = ARXIV:0907.4658;%%.

\bibitem{Bergshoeff:2010mf}
E.~A. Bergshoeff, O.~Hohm, J.~Rosseel, E.~Sezgin, and P.~K. Townsend, ``{More
  on Massive 3D Supergravity},''
  \href{http://dx.doi.org/10.1088/0264-9381/28/1/015002}{{\em Class. Quant.
  Grav.} {\bf 28} (2011)  015002},
\href{http://arxiv.org/abs/1005.3952}{{\tt arXiv:1005.3952 [hep-th]}}.
%%CITATION = ARXIV:1005.3952;%%.

\bibitem{Bergshoeff:2010iy}
E.~A. Bergshoeff, O.~Hohm, J.~Rosseel, E.~Sezgin, and P.~K. Townsend, ``{On
  Critical Massive (Super)Gravity in adS3},''
  \href{http://dx.doi.org/10.1088/1742-6596/314/1/012009}{{\em J. Phys. Conf.
  Ser.} {\bf 314} (2011)  012009},
\href{http://arxiv.org/abs/1011.1153}{{\tt arXiv:1011.1153 [hep-th]}}.
%%CITATION = ARXIV:1011.1153;%%.

\bibitem{Alkac:2014hwa}
G.~Alkac, L.~Basanisi, E.~A. Bergshoeff, M.~Ozkan, and E.~Sezgin, ``{Massive $
  \mathcal{N} $ = 2 supergravity in three dimensions},''
  \href{http://dx.doi.org/10.1007/JHEP02(2015)125}{{\em JHEP} {\bf 02} (2015)
  125},
\href{http://arxiv.org/abs/1412.3118}{{\tt arXiv:1412.3118 [hep-th]}}.
%%CITATION = ARXIV:1412.3118;%%.

\bibitem{Kuzenko:2011rd}
S.~M. Kuzenko and G.~Tartaglino-Mazzucchelli, ``{Three-dimensional N=2 (AdS)
  supergravity and associated supercurrents},''
  \href{http://dx.doi.org/10.1007/JHEP12(2011)052}{{\em JHEP} {\bf 12} (2011)
  052},
\href{http://arxiv.org/abs/1109.0496}{{\tt arXiv:1109.0496 [hep-th]}}.
%%CITATION = ARXIV:1109.0496;%%.

\bibitem{Kuzenko:2013uya}
S.~M. Kuzenko, U.~Lindstrom, M.~Rocek, I.~Sachs, and
  G.~Tartaglino-Mazzucchelli, ``{Three-dimensional $\mathcal{N} =$ 2
  supergravity theories: From superspace to components},''
  \href{http://dx.doi.org/10.1103/PhysRevD.89.085028}{{\em Phys. Rev.} {\bf
  D89} (2014) no.~8, 085028},
\href{http://arxiv.org/abs/1312.4267}{{\tt arXiv:1312.4267 [hep-th]}}.
%%CITATION = ARXIV:1312.4267;%%.

\bibitem{Kuzenko:2015jda}
S.~M. Kuzenko, J.~Novak, and G.~Tartaglino-Mazzucchelli, ``{Higher derivative
  couplings and massive supergravity in three dimensions},''
  \href{http://dx.doi.org/10.1007/JHEP09(2015)081}{{\em JHEP} {\bf 09} (2015)
  081},
\href{http://arxiv.org/abs/1506.09063}{{\tt arXiv:1506.09063 [hep-th]}}.
%%CITATION = ARXIV:1506.09063;%%.

\bibitem{Deger:2013yla}
N.~S. Deger, A.~Kaya, H.~Samtleben, and E.~Sezgin, ``{Supersymmetric Warped AdS
  in Extended Topologically Massive Supergravity},''
  \href{http://dx.doi.org/10.1016/j.nuclphysb.2014.04.011}{{\em Nucl. Phys.}
  {\bf B884} (2014)  106--124},
\href{http://arxiv.org/abs/1311.4583}{{\tt arXiv:1311.4583 [hep-th]}}.
%%CITATION = ARXIV:1311.4583;%%.

\bibitem{Alkac:2015lma}
G.~Alkac, L.~Basanisi, E.~A. Bergshoeff, D.~O. Devecioglu, and M.~Ozkan,
  ``{Supersymmetric backgrounds and black holes in $
  \mathcal{N}=\left(1,\;1\right) $ cosmological new massive supergravity},''
  \href{http://dx.doi.org/10.1007/JHEP10(2015)141}{{\em JHEP} {\bf 10} (2015)
  141},
\href{http://arxiv.org/abs/1507.06928}{{\tt arXiv:1507.06928 [hep-th]}}.
%%CITATION = ARXIV:1507.06928;%%.

\bibitem{Deger:2016vrn}
N.~S. Deger and G.~Moutsopoulos, ``{Supersymmetric solutions of $N=(2,0)$
  Topologically Massive Supergravity},''
  \href{http://dx.doi.org/10.1088/0264-9381/33/15/155006}{{\em Class. Quant.
  Grav.} {\bf 33} (2016) no.~15, 155006},
\href{http://arxiv.org/abs/1602.07263}{{\tt arXiv:1602.07263 [hep-th]}}.
%%CITATION = ARXIV:1602.07263;%%.

\bibitem{Rocek:1985bk}
M.~Rocek and P.~van Nieuwenhuizen, ``{$N \geq 2$ supersymmetric Chern-Simons
  terms as $d = 3$ extended conformal supergravity},''
\href{http://dx.doi.org/10.1088/0264-9381/3/1/007}{{\em Class. Quant. Grav.}
  {\bf 3} (1986)  43}.
%%CITATION = CQGRD,3,43;%%.

\bibitem{Nishino:1991sr}
H.~Nishino and S.~J. Gates, Jr., ``{Chern-Simons theories with supersymmetries
  in three-dimensions},''
\href{http://dx.doi.org/10.1142/S0217751X93001363}{{\em Int. J. Mod. Phys.}
  {\bf A8} (1993)  3371--3422}.
%%CITATION = IMPAE,A8,3371;%%.

\bibitem{Cecotti:2010dg}
E.~Bergshoeff, S.~Cecotti, H.~Samtleben, and E.~Sezgin, ``{Superconformal Sigma
  Models in Three Dimensions},''
  \href{http://dx.doi.org/10.1016/j.nuclphysb.2010.04.023}{{\em Nucl. Phys.}
  {\bf B838} (2010)  266--297},
\href{http://arxiv.org/abs/1002.4411}{{\tt arXiv:1002.4411 [hep-th]}}.
%%CITATION = ARXIV:1002.4411;%%.

\bibitem{Becker:2009mk}
M.~Becker, P.~Bruillard, and S.~Downes, ``{Chiral Supergravity},''
  \href{http://dx.doi.org/10.1088/1126-6708/2009/10/004}{{\em JHEP} {\bf 10}
  (2009)  004},
\href{http://arxiv.org/abs/0906.4822}{{\tt arXiv:0906.4822 [hep-th]}}.
%%CITATION = ARXIV:0906.4822;%%.

\bibitem{Grumiller:2009mw}
D.~Grumiller and I.~Sachs, ``{AdS (3) / LCFT (2) -- Correlators in Cosmological
  Topologically Massive Gravity},''
  \href{http://dx.doi.org/10.1007/JHEP03(2010)012}{{\em JHEP} {\bf 03} (2010)
  012},
\href{http://arxiv.org/abs/0910.5241}{{\tt arXiv:0910.5241 [hep-th]}}.
%%CITATION = ARXIV:0910.5241;%%.

\bibitem{Grumiller:2009sn}
D.~Grumiller and O.~Hohm, ``{AdS(3)/LCFT(2): Correlators in New Massive
  Gravity},'' \href{http://dx.doi.org/10.1016/j.physletb.2010.02.065}{{\em
  Phys. Lett.} {\bf B686} (2010)  264--267},
\href{http://arxiv.org/abs/0911.4274}{{\tt arXiv:0911.4274 [hep-th]}}.
%%CITATION = ARXIV:0911.4274;%%.

\bibitem{Grumiller:2010tj}
D.~Grumiller, N.~Johansson, and T.~Zojer, ``{Short-cut to new anomalies in
  gravity duals to logarithmic conformal field theories},''
  \href{http://dx.doi.org/10.1007/JHEP01(2011)090}{{\em JHEP} {\bf 01} (2011)
  090},
\href{http://arxiv.org/abs/1010.4449}{{\tt arXiv:1010.4449 [hep-th]}}.
%%CITATION = ARXIV:1010.4449;%%.

\bibitem{Khorrami:1998kw}
M.~Khorrami, A.~Aghamohammadi, and A.~M. Ghezelbash, ``{Logarithmic N=1
  superconformal field theories},''
  \href{http://dx.doi.org/10.1016/S0370-2693(98)01029-6}{{\em Phys. Lett.} {\bf
  B439} (1998)  283--288},
\href{http://arxiv.org/abs/hep-th/9803071}{{\tt arXiv:hep-th/9803071
  [hep-th]}}.
%%CITATION = HEP-TH/9803071;%%.

\bibitem{Drichel:2010eb}
D.~Drichel and M.~Flohr, ``{Correlation Functions in N=3 Superconformal
  Theory},''
\href{http://arxiv.org/abs/1006.3346}{{\tt arXiv:1006.3346 [hep-th]}}.
%%CITATION = ARXIV:1006.3346;%%.

\bibitem{Pearce:2013bea}
P.~A. Pearce, J.~Rasmussen, and E.~Tartaglia, ``{Logarithmic Superconformal
  Minimal Models},''
  \href{http://dx.doi.org/10.1088/1742-5468/2014/05/P05001}{{\em J. Stat.
  Mech.} {\bf 1405} (2014)  P05001},
\href{http://arxiv.org/abs/1312.6763}{{\tt arXiv:1312.6763 [hep-th]}}.
%%CITATION = ARXIV:1312.6763;%%.

\bibitem{Grumiller:2009dx}
D.~Grumiller and P.~van Nieuwenhuizen, ``{Holographic counterterms from local
  supersymmetry without boundary conditions},''
  \href{http://dx.doi.org/10.1016/j.physletb.2009.11.022}{{\em Phys. Lett.}
  {\bf B682} (2010)  462--465},
\href{http://arxiv.org/abs/0908.3486}{{\tt arXiv:0908.3486 [hep-th]}}.
%%CITATION = ARXIV:0908.3486;%%.

\end{thebibliography}
%\bibliographystyle{newutphys}

\providecommand{\href}[2]{#2}\begingroup\raggedright\endgroup

\end{document}